\documentclass[11pt,letterpaper]{article}

\usepackage{amssymb}
\usepackage{amsmath}
\usepackage{color}
\usepackage[pdftex]{graphicx}
\usepackage{url} 
\usepackage[margin=1in]{geometry}

\newtheorem{cor}{Corollary}

\newtheorem{lem}{Lemma}

\newtheorem{rem}{Remark}
\newtheorem{theo}{Theorem}

\begin{document}

\title{\textsc{Kuhn's Theorem for Games of the Extensive Form with Unawareness\thanks{We thank two reviewers and the associate editor for detailed suggestions. We also thank Aviad Heifetz and Martin Meier for many discussions about games of the extensive form with unawareness and participants in LOFT 2018 and at a 2025 NUS Theory Seminar for helpful comments. Financial support through NSF SES-0647811 and ARO Contract W911NF2210282 is gratefully acknowledged.}}}

\author{Ki Vin Foo\thanks{Department of Economics, National University of Singapore. Email: kivin.foo@u.nus.edu} \and Burkhard C. Schipper\thanks{Department of Economics, University of California, Davis. Email: bcschipper@ucdavis.edu}}

\date{April 16, 2026}

\maketitle

\begin{abstract} We extend Kuhn's Theorem to games of the extensive form with unawareness. We prove that if a game of the extensive form with unawareness has perfect recall, then for each mixed strategy there is an equivalent behavior strategy. We show that the converse does not hold under unawareness without restricting the evolution of the player's awareness to constant awareness along paths of play. Both directions of Kuhn's Theorem for games of the extensive form with unawareness require a condition complementary to perfect recall that rules out falsely believing in some events when the player is unaware of the actual past events. 
\bigskip


\bigskip

\noindent \textbf{JEL-Classifications: } C72, D83.
\end{abstract}

\thispagestyle{empty}

\pagenumbering{empty}

\renewcommand{\baselinestretch}{1.2}

\small\normalsize

\pagenumbering{arabic}

\newpage

\section{Introduction}

In games of the extensive form there are two notions of mixed strategies. First, a player can ex ante mix over her strategies, using a so called mixed strategy. Second, at each information set a player can independently mix interim over actions at that information set, using a so called behavior strategy. Kuhn's Theorem states that a game of the extensive form has perfect recall if and only if for each mixed strategy there is a behavior strategy that is equivalent in terms of the probabilities of reaching nodes in the game tree (Kuhn, 1953). Since equilibrium may not be feasible in games with imperfect recall and the restriction to behavior strategies has been extremely useful both in theory development and applications, nearly the entire game theoretic and applied literature invokes the perfect recall assumption\footnote{See Piccione and Rubinstein (1997) and the contributions to the 1997 special issue in Games and Economic Behavior for exceptions.} and often invokes at least one implication of Kuhn's Theorem. In that regard, it is fair to say that Kuhn's Theorem is one of the most fundamental theorems on games of the extensive form. 

More recently, the theory of games of the extensive form has been extended to unawareness (e.g., Halpern and R\^{e}go, 2014, Heifetz, Meier, and Schipper, 2013, Grant and Quiggin, 2013, Feinberg, 2021, R\^{e}go and Halpern, 2012, Ozbay, 2007; see Schipper, 2014, for a brief review). Unawareness refers to the lack of conception rather than just the lack of information. To advance this theory, it is instrumental to extend also the most fundamental results for standard games to games with unawareness or explore how unawareness limits such extensions. In this paper, we extend Kuhn's Theorem to games of the extensive form with unawareness. The result is useful for further theory development and applications. For instance, Kuhn's Theorem for games with unawareness has been applied by Schipper (2021) to define and to show the existence of extensive-form rationalizable self-confirming equilibria for games with unawareness, an equilibrium notion for games with unawareness that can be understood as a result of a process of both discovering features of the game and learning about the behavior of opponents. Our extension of Kuhn's Theorem to games of the extensive form with unawareness is also interesting from a purely game theoretic perspective as it allows us to learn how far we can meaningfully generalize the perfect recall assumption and notions of mixed and behavior strategies beyond standard games. 

Extending the perfect recall assumption and Kuhn's Theorem to games of the extensive form with unawareness is not straightforward. According to Kuhn (1953, p. 213), perfect recall ``is equivalent to the assertion that each player is allowed by the rules of the game to remember everything she knew at previous moves and all of her choices at those moves.''\footnote{See also Ritzberger (1999) for a characterization of perfect recall and the discussion of perfect recall in Ritzberger (2002).} In games that allow for unawareness, it must also mean that players do not lose awareness of actions or moves of nature that they previously were able to consider, a feature that is mute in standard games. As we will discuss in the paper, this poses a challenge for obtaining the converse of Kuhn's Theorem, namely that the equivalence of mixed and behavior strategies implies perfect recall. Even more fundamentally, while perfect recall means that players are able to remember their previous experiences and actions, it does not rule out that they ``make up'' past events leading to an information set if they are otherwise unable to rationalize having arrived at an information set due to the lack of awareness of some features of the game. As we will demonstrate in the main text, this poses a challenge for both directions of Kuhn's Theorem and requires us to introduce a condition that is complementary to perfect recall and rules out such ``delusions''.  

There are further subtleties in games with unawareness that pose a challenge for the extension of Kuhn's Theorem. Games of the extensive form with unawareness feature a forest of game trees rather than just one tree. The information set at a node in one tree may actually reside in a less expressive tree, signifying the fact that the player to whom this information set belongs to is unaware of something. These obvious differences in the formalism to standard games of the extensive form make it non-trivial to define perfect recall as it must capture recall in the face of the player's evolution of awareness. The notion of mixed strategy is less compelling in games with unawareness. A player may not be aware of all actions ex ante. Thus, she may not be able to ex ante conceive of all possible strategies and consequently may be unable to mix over them. Yet, given her awareness ex ante, she conceives of some \emph{partial} strategies that she is able to mix over. As the play unfolds, the players' awareness may change and information sets in more expressive trees become relevant. The player will now be able to conceive of a larger set of strategies. It begs now the question whether for each (partial) mixed strategy there is an equivalent (partial) behavior strategy. However, what is a meaningful notion of two strategies being equivalent to each other? Games of the extensive form with unawareness allow the game theorist to model differing players' subjective conceptions of the game during the play. Since (partial) strategies are objects of players' beliefs, the notion of a strategy \emph{reaching} a node or an information set is also subjective. A player may believe that a partial strategy profile reaches a particular node or a particular information set even though this node or information set cannot ``objectively'' \emph{occur} with this partial strategy profile. We formalize these two notions of a strategy being consistent with a node or information set, show their relationship, and extend Kuhn's Theorem to both of these two notions in games of the extensive form with unawareness. 

We summarize our formal results: We show that if the game of the extensive form with unawareness satisfies perfect recall, then for every mixed strategy there exists an equivalent behavior strategy. We refer to this statement as Kuhn's Theorem. We show by example that Kuhn's Theorem crucially depends on strengthening the condition of Generalized Reflexivity (Heifetz, Meier, and Schipper, 2013) in order to rule out delusions w.r.t. actions that have been taken in the past. 
 
While major textbooks\footnote{Well known textbooks like Osborne and Rubinstein (1994) and Maschler, Solan, and Zamir (2013) state and prove one direction. Even Kuhn's (2002) own lecture notes just proves one direction. A notable exception is Ritzberger (2002) who presents the most comprehensive treatment in the literature and states and proves both directions.}  and even the seminal paper by Selten (1975) just state that perfect recall implies the equivalence of mixed and behavior strategies for standard games (without unawareness), Kuhn (1953) also proved the converse, namely, if for every mixed strategy there is an equivalent behavior strategy, then the game must have perfect recall. This direction is less commonly considered, but a complete analysis of perfect recall in games with unawareness should also provide an answer w.r.t. the converse. We show by example that our strengthening of generalized reflexivity is also necessary for the converse. We present another example showing that nevertheless the converse does not obtain for games with unawareness in general. Then, we find a new condition restricting the evolution of awareness that remedies the aforementioned failure of the converse and show that under this additional condition, in what we term \textit{restricted} games of the extensive form with unawareness, we obtain a characterization of the equivalence of mixed and behavior strategies with perfect recall. The condition under which we are able to obtain a characterization restricts the evolution of awareness to constant awareness along paths of play in each tree. While it still allows for asymmetric awareness, it is a strong condition. We show with the help of another example that there is no hope of weakening the condition if one desires the converse. We also discuss that weakening the condition of perfect recall does not lead to a meaningful converse because it would have to permit losing awareness of previously considered actions. 

Finally, we extend Kuhn's Theorem to the second notion of a strategy profile being consistent with a node, the notion of a node ``objectively'' occurring with a strategy profile rather than ``subjectively'' being reached by a strategy profile. We also extend Kuhn's Theorem to $T$-partial games, where a $T$-partial game consists of a game of the extensive form with unawareness for which the join of trees is $T$ and all relevant less expressive trees. Such $T$-partial games model the strategic situation from the point of view of a player whose awareness level is given by whatever can be described by the tree $T$. 

We conclude with a discussion of the notions of perfect recall and strategies in games with unawareness. In particular, we speculate about whether players could take actions of which they themselves are unaware of and point to a connection to what has been called situation awareness in military science and psychology. 

We hope that our extension of Kuhn's Theorem further facilitates the growing literature on unawareness. Various applications of games of the extensive form with unawareness have already been developed in the literature. For instance, verifiable disclosure and the failure of unraveling of information are analyzed by Heifetz, Meier, and Schipper (2021); see Li and Schipper (2025) for an experiment. Schipper and Woo (2019) study political awareness in political campaigning. Filiz-Ozbay (2012) analyzes disclosure before insurance contracts, Francetich and Schipper (2026) analyze the incentives of an agent for raising the awareness of a principal before screening by the latter, and Pram and Schipper (2025) introduce dynamic elaboration VCG mechanisms for implementation of efficient outcomes at the pooled awareness level in conditional dominant strategies. We have now all game theoretic tools ready to revisit interesting problems in game theory and economics, and explore how the presence of unawareness may change predictions; see Schipper (2025) for a bibliography of the growing literature. 

The paper is organized as follows: The next section spells out in detail games of the extensive form with unawareness. Section~\ref{strategies} introduces various notions of strategies. The formal results are contained in Section~\ref{KT_section}. Section~\ref{discussion} concludes with a discussion.

\section{Games of the Extensive form with Unawareness\label{model}}

In this section, we outline games of the extensive form with unawareness \`{a} la Heifetz, Meier, and Schipper (2013).\footnote{Although there are differences in the formalism between various approaches to games of the extensive form with unawareness (Halpern and R\^{e}go, 2014, Heifetz, Meier, and Schipper, 2013, Grant and Quiggin, 2013, Feinberg, 2021, R\^{e}go and Halpern, 2012, Ozbay, 2007; see Schipper, 2014, for a brief review), all approaches model unawareness that is consistent with the paradigm of ``propositional awareness'' as in Fagin and Halpern (1988) or Heifetz, Meier, and Schipper (2006). We use here the approach by Heifetz, Meier, and Schipper (2013) (also used by Schipper and Woo, 2019, Heifetz, Meier, and Schipper, 2021, Schipper, 2021, Schipper, 2026, Francetich and Schipper, 2026, Li and Schipper, 2025, and Pram and Schipper, 2025) because information sets in their approach can be interpreted as states of the mind of a player at a node. In their approach, information sets model both information and awareness (i.e., ``explicit'' information) rather than just ``information if the player were aware of it'' (i.e., ``implicit'' information). It also avoids having to define a separate awareness correspondence that for each node specifies which nodes the player is aware of.} To define a game of the extensive form with unawareness $\Gamma$, consider first, as a building block, a finite game of perfect information with possibly simultaneous moves. That is, we allow for simultaneous moves as in Dubey and Kaneko, 1984, or in Osborne and Rubinstein, 1994, Chapter 6.3.2. The major purpose of this tree is to outline all physical moves. There is a finite set of players $I$ and possibly a special player ``nature'' with index $0$. We denote by $I^0$ the set of players including nature. Further, there is a nonempty finite set of ``decision'' nodes $\bar{D}$ and a player correspondence $P: \bar{D} \longrightarrow 2^{I^0} \setminus \{\emptyset\}$ that assigns to each node $n \in \bar{D}$, a nonempty set of ``active'' players $P(n) \subseteq I^0$. For every decision node $n \in \bar{D}$ and player $i \in P(n)$ who moves at that decision node, there is a nonempty finite set of actions $A_i(n)$. Moreover, there is a set of terminal nodes $\bar{Z}$. Each terminal node $z \in \bar{Z}$ is associated with a vector of payoffs $(u_i(z))_{i \in I}$, one for each player $i \in I$. We require that nodes in $\bar{N}$ are partially ordered by a precedence relation $\bar{\lessdot}$ with which $(\bar{N}, \bar{\lessdot})$ forms an arborescence (that is, the predecessors of each node in $\bar{N}$ are totally ordered by $\bar{\lessdot}$). This means that nodes in $\bar{N} := \bar{D} \cup \bar{Z}$ constitute a directed tree denoted by $\bar{T}$ and there is a unique node in $\bar{N}$ with no predecessors (i.e., the root of the tree). Finally, for each decision node $n \in \bar{D}$ there is a bijection $\psi_n$ between the action profiles $\prod_{i \in P(n)} A_i(n)$ at $n$ and $n$'s immediate successors. Any terminal node in $\bar{Z}$ has no successors.

Note that so far we treat nature like any other player except that at terminal nodes we do not assign payoffs to nature.\footnote{Alternatively, we could assign at every terminal node the same payoff to nature.} We do not need to require that nature moves first or that nature moves according to a pre-specified probability distribution (although these assumptions can be imposed in our framework).

Consider now a finite join-semilattice $\mathbf{T}$ of subtrees of $\bar{T}$.\footnote{A join semi-lattice is a partially ordered set in which each pair of elements has a join, i.e., a least upper bound. Requiring the forest of trees to form a join-semilattice implies there is an objective set of rules governing the strategic interaction. Moreover, no matter which action a player may become aware of during the play, there is a representation that captures all actions that she is aware of.} A subtree $T$ is defined by a subset of nodes $N \subseteq \bar{N}$ with $N \cap \bar{D} \neq \emptyset$, along with $\lessdot$ the restriction of $\bar{\lessdot}$ to $N$, where $(N, \lessdot)$ is also a tree. Two subtrees $T', T'' \in \mathbf{T}$ are ordered, written
\begin{equation*}
T' \preceq T''
\end{equation*}
if the nodes of $T'$ constitute a subset of the nodes of $T''$.

Within the family $\mathbf{T}$ of subtrees of $\bar{T}$, some nodes $n$ appear in several trees $T \in \mathbf{T}$. In what follows, we will need to designate explicitly appearances of such nodes $n$ in different trees as distinct objects.\footnote{It turns out that it is much easier to talk about copies of nodes in a less expressive tree than copies of histories in a less expressive tree. This is the main reason for why we make use of the ``older'' definition of games of extensive form based on nodes (Kuhn, 1953, Selten, 1975) rather than the more recent notion based on histories (Osborne and Rubinstein, 1994).} To this effect, in each tree $T \in \mathbf{T}$ label by $n_{T}$ the copy in $T$ of the node $n \in \bar{N}$ whenever the copy of $n$ is part of the tree $T$, with the requirement that if the profile of actions $a_{n} \in A(n)$ leads from $n$ to $n'$, then $a_{n_T}$ leads also from the copy $n_{T}$ to the copy $n_{T}'$.\footnote{One can also consider each subtree $T \in \mathbf{T}$ to be a reduced version of $\bar{T}$, and node $n_T$ in tree $T$ is a copy of node $n$ in $\bar{T}$ if the sequence of actions leading up to $n_T$ before the reduction to $T$ is identical to that of $n$ in $\bar{T}$.} For any $T, T', T'' \in \mathbf{T}$ with $T \preceq T' \preceq T''$ such that $n \in T''$, $n_{T'}$ is the copy of $n$ in the tree $T'$, $n_T$ is the copy of $n$ in the tree $T$, and $(n_{T'})_{T}$ is the copy of $n_{T'}$ in the tree $T$, we require that ``nodes commute'', $n_T = (n_{T'})_T$. For any $T \in \mathbf{T}$ and any $n \in T$, we let $n_T := n$ (i.e., the copy of $n \in T$ in $T$ is $n$ itself).

We require the following properties:
\begin{itemize}
\item[0.] For any $n, n', n'' \in \bar{T}$ such that $n \ \bar{\lessdot} \ n' \ \bar{\lessdot} \ n''$, if there exists $n_T, n''_T \in T$ then there is $n'_T \in T$. That is, we do not allow decision nodes `in-between' two nodes in the objective tree to be removed in a subtree unless one of the two nodes is also removed.

\item[1.] All the terminal nodes in each tree $T \in \mathbf{T}$ are in $\bar{Z}$. That is, we don't create ``new'' terminal nodes.

\item[2.] For every tree $T \in \mathbf{T}$, every node $n \in T$, and every active player $i \in P(n)$ there exists a nonempty subset of actions $A^T_i(n) \subseteq A_i(n)$ such that $\psi_{n}$ maps the action profiles $A^T(n) = \prod_{i \in P(n)} A^T_i(n)$ bijectively onto $n$'s successors in $T$.

\item[3.] For any tree $T \in \mathbf{T}$, if for two decision nodes $n, n' \in T$ with $i \in P(n) \cap P(n')$ it is the case that $A_i(n) \cap A_i(n') \neq \emptyset $, then $A_i(n) = A_i(n')$.
\end{itemize}

\begin{figure}[!h]
\caption{Property 0\label{property0}}
\begin{center}
\includegraphics[scale=.25]{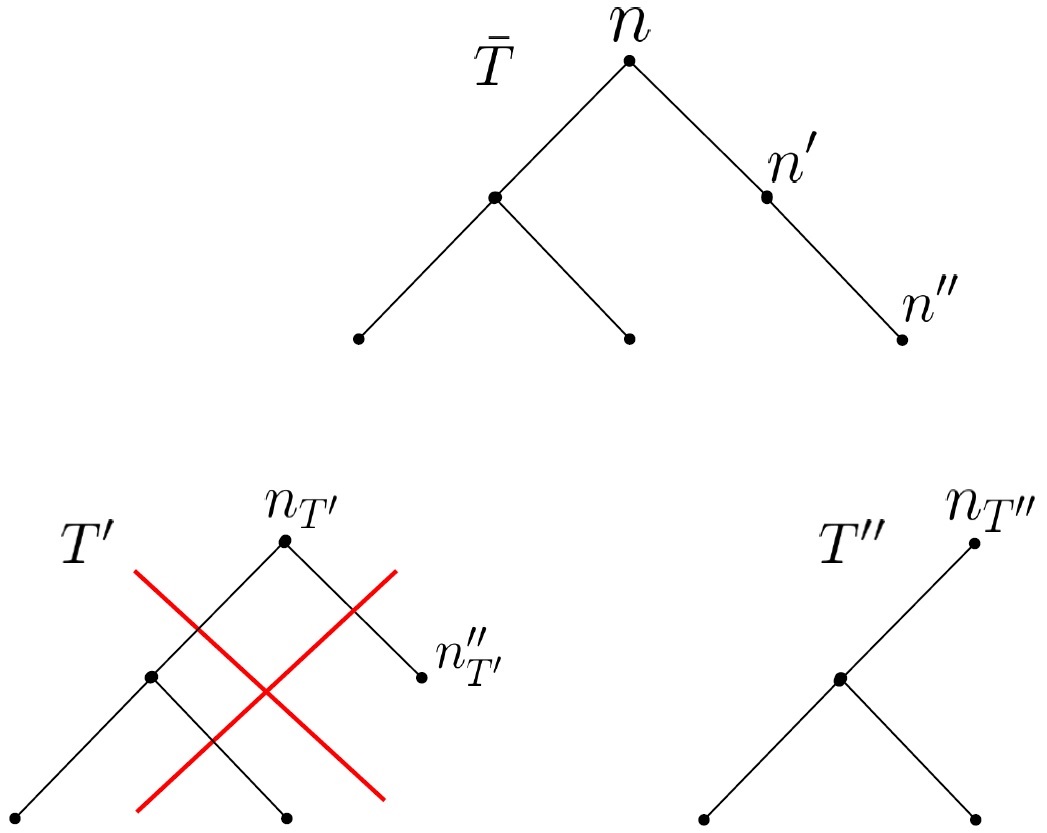}
\end{center}
\end{figure}
We illustrate Property 0 in Figure~\ref{property0}. Suppose that the tree modeling all physical moves is given by $\bar{T}$. Then $T'$ is a subtree that does not satisfy Property 0. This is because nodes $n, n', n''$ are in $\bar{T}$ and $n \ \bar{\lessdot} \ n' \ \bar{\lessdot} \ n''$, however in $T'$ there are copies of nodes $n$ and $n''$ (that is, $n_{T'}$ and $n''_{T'}$) but there is no copy of $n'$ in $T'$. That is, the decision node $n'$ between $n$ and $n''$ in $\bar{T}$ is not in $T'$ when $n_{T'}$ and $n''_{T'}$ exist and that is not allowed under Property 0. Subtree $T''$ on the other hand satisfies Property 0, because even though there is no copy of $n'$ in $T''$, there is also no copy of $n''$ in $T''$.

\begin{figure}[!h]
\caption{Property 1\label{property1}}
\begin{center}
\includegraphics[scale=.25]{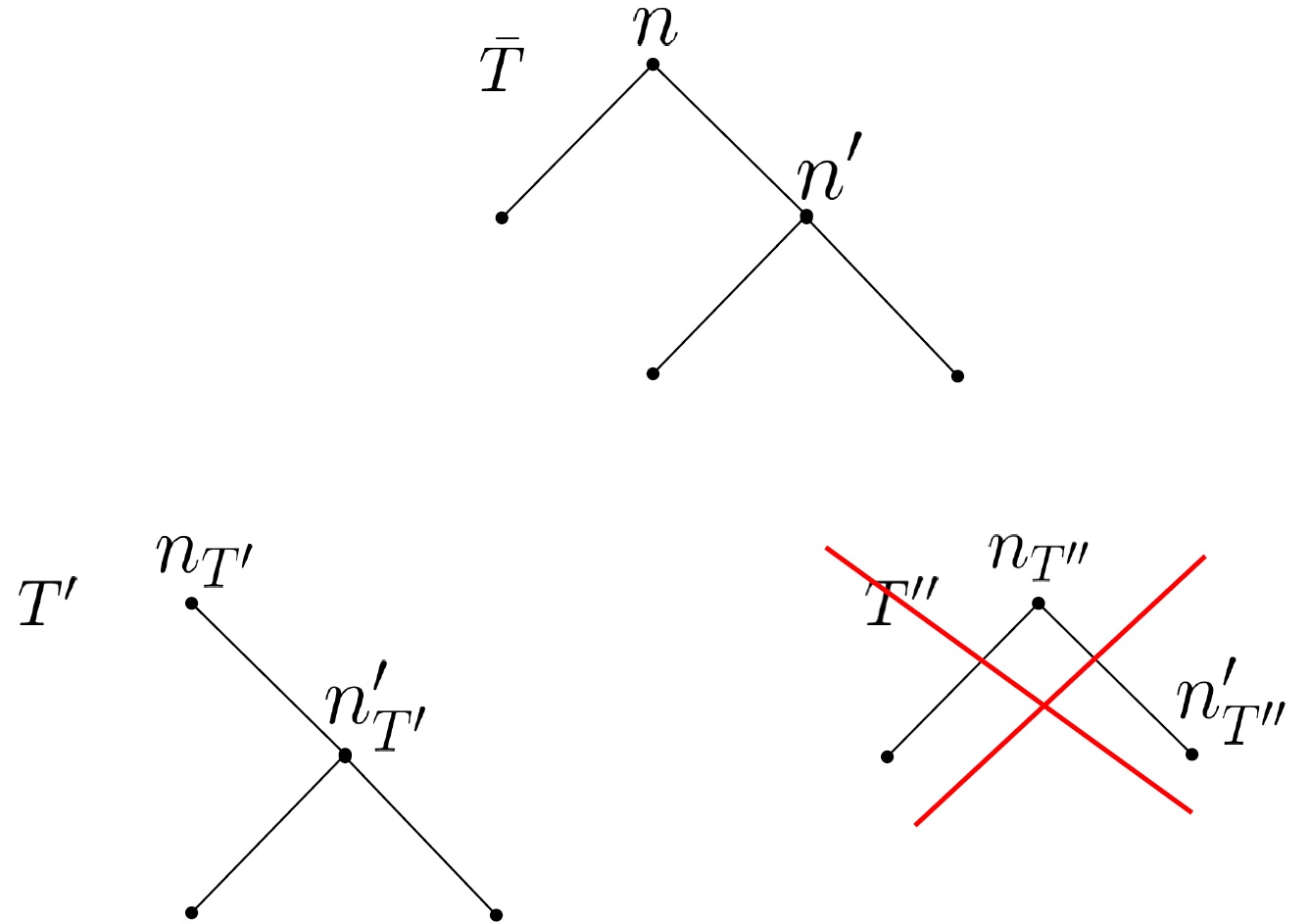}
\end{center}
\end{figure}

Next, we illustrate Property 1 in Figure~\ref{property1}. Suppose that the tree the modeling all physical moves is given by $\bar{T}$. Then $T'$ is a subtree satisfying Property 1. In contrast, tree $T''$ does not satisfy Property 1 because it contains a new terminal node $n'_{T''}$. After cutting branches from tree $\bar{T}$ to create tree $T''$, there is now a node $n'_{T''}$ in $T''$ that was not a terminal node in the original tree $\bar{T}$, i.e., $n'$ is not a terminal node in $\bar{T}$.

Denote by $\mathbf{D}$ the union of all decision nodes in all trees $T \in \mathbf{T}$, by $\mathbf{Z}$ the union of terminal nodes in all trees $T \in \mathbf{T}$, and by $\mathbf{N} = \mathbf{D} \cup \mathbf{Z}$. Recall that copies $n_{T}$ of a given node $n$ in different subtrees $T$ are now treated distinct from one another, so that $\mathbf{N}$ is a disjoint union of sets of nodes.

In what follows, when referring to a node in $\mathbf{N}$ we will typically avoid the subscript indicating the tree $T$ for which $n \in T$ when no confusion arises. For a node $n \in \mathbf{N}$ we denote by $T_{n}$ the tree containing $n$.\footnote{Bold capitalized letters refer to sets of elements across trees.}

Denote by $N^{T}$ the set of nodes in the tree $T \in \mathbf{T}$. Similarly, denote by $D_i^T$ the set of decision nodes in which player $i$ is active in the tree $T \in \mathbf{T}$. Moreover, denote by $Z^T$ the set of terminal nodes in the tree $T \in \mathbf{T}$. Finally, we let $\mathbf{D}_i$ denote the set of player $i$'s decision nodes over all trees in $\mathbf{T}$. In the case of nature, $\mathbf{D}_0$ would be all nodes at which nature moves.

In games of the extensive form with unawareness, information sets model both information and awareness. At decision node $n$ of player $i$ in the tree $T_{n} \in \mathbf{T}$, the player may conceive the feasible paths to be described by a different (i.e., less expressive) tree $T' \in \mathbf{T}$. In such a case, her information set will be a subset of $T'$ rather than of $T_{n}$ and $n$ will not be contained in the player's information set at $n$.

Formally, for each node $n \in \mathbf{N}$, define for each active player $i \in P(n) \setminus \{0\}$ a nonempty information set $h_i(n)$ with the following properties:\footnote{We keep the numbering consistent with Heifetz, Meier, and Schipper (2013).}

\begin{itemize}
\item[I2] Introspection: If $n' \in h_i(n)$, then $h_i(n') = h_i(n)$.

\item[I3] No divining of currently unimaginable paths, no expectation to forget currently conceivable paths: If $n' \in h_i(n) \subseteq T'$ (where $T' \in \mathbf{T}$ is a tree) and there is a path $n', \dots , n'' \in T'$ such that $i \in P(n') \cap P(n'')$, then $h_i(n'') \subseteq T'$.

\item[I4] No imaginary actions: If $n' \in h_i(n)$, then $A_i(n') \subseteq A_i(n)$.

\item[I5] Distinct action names in disjoint information sets: For a subtree $T \in \mathbf{T}$, if there are decision nodes $n, n' \in T \cap \mathbf{D}$ with $A_i(n) = A_i(n')$, then $h_i(n') = h_i(n)$.

\item[U0] Confined awareness: If $n \in T$ and $i \in P(n)$, then there is $T' \in \mathbf{T}$, $T' \preceq T$ such that $h_i(n) \subseteq T'$.

\item[U1] Nondelusion: If $n \in T$, $h_i(n) \subseteq T'$, then there is $n_{T'}$ such that $n_{T'} \in h_i(n)$.\footnote{In past work a weaker condition Generalized Reflexivity was used instead. It will not be sufficient for our purposes in this paper. We elaborate in Section~\ref{KT_section}.}
\end{itemize}

\begin{figure}[h!]
\caption{Property U0}
\label{properties}
\begin{center}
\includegraphics[scale=0.25]{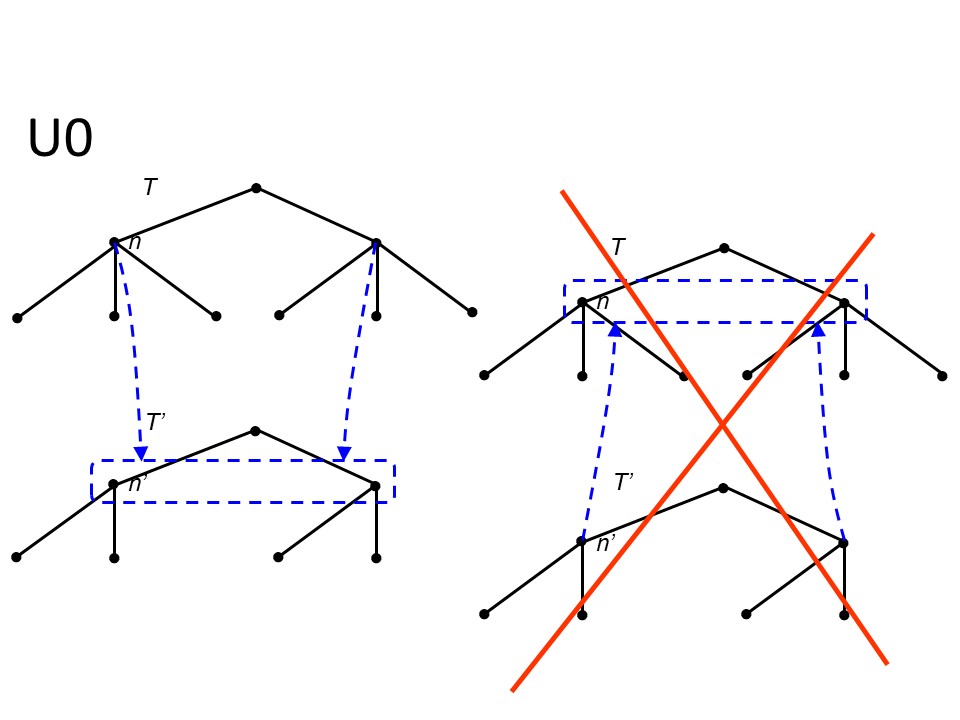}
\end{center}
\end{figure}

Properties I2, I4, and I5 are standard for games of the extensive form. Property I3 confines at each information set the player's \emph{anticipation} of her future view of the game to the view she currently holds (even if, as a matter of fact, this view is about to be shattered as the game evolves); see Heifetz, Meier, and Schipper (2013) for further discussion and illustration. Properties U0 and U1 generalize standard properties of games of the extensive form to our generalized setting with unawareness. Confined awareness (U0) says that at each point in the game, the player has a well-defined view of the game. If this condition would be violated, i.e., there is an information of the player that would contain nodes of different trees, then the player would be uncertain about the game tree. But this could be modeled by a more comprehensive tree. Figure~\ref{properties} illustrates with an example and a counterexample to U0. Nondelusion (U1) is a strengthening of a condition Generalized reflexivity in Heifetz, Meier, and Schipper (2013). We will discuss this condition after Theorem~\ref{Kuhn}. 

Central to our extension of Kuhn's Theorem is the condition of Perfect recall (I6). 

\begin{itemize}
\item[I6] Perfect recall: Suppose that player $i$ is active at two distinct nodes $n_{1}$ and $n_{k}$, and there is a path $n_{1}, n_{2}, ..., n_{k}$ such that at $n_{1}$ player $i$ takes the action $a_{i}$. If $n^{\prime} \in h _{i}\left( n_{k}\right)$, then there exists a node $n_{1}^{\prime}\neq n^{\prime }$ and a path $n_{1}^{\prime}, n_{2}^{\prime }, ..., n_{\ell}^{\prime } = n^{\prime }$ such that $h_{i}\left( n_{1}^{\prime}\right) = h_{i} \left( n_{1}\right)$ and at $n_{1}^{\prime }$ player $i$ takes the action $a_{i}$.
\end{itemize}

\begin{figure}[h!]
\caption{Perfect Recall Example\label{I6_1}}
\begin{center}
\includegraphics[scale=.3]{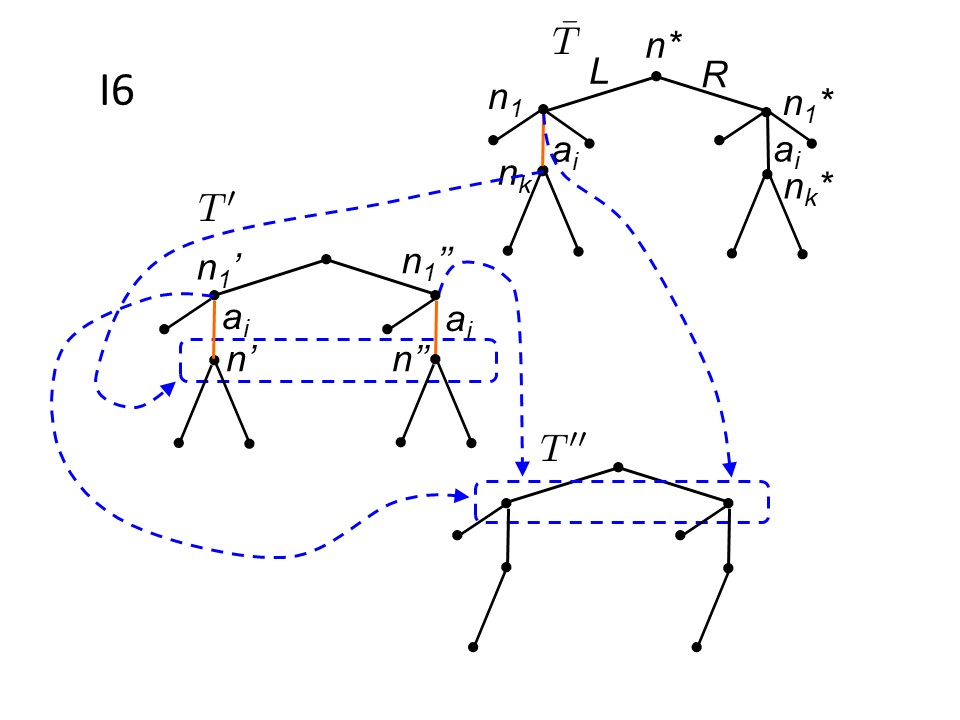}
\end{center}
\end{figure}

\begin{figure}[h!]
\caption{Perfect Recall Counterexamples\label{I6_2}}
\begin{center}
\includegraphics[scale=.3]{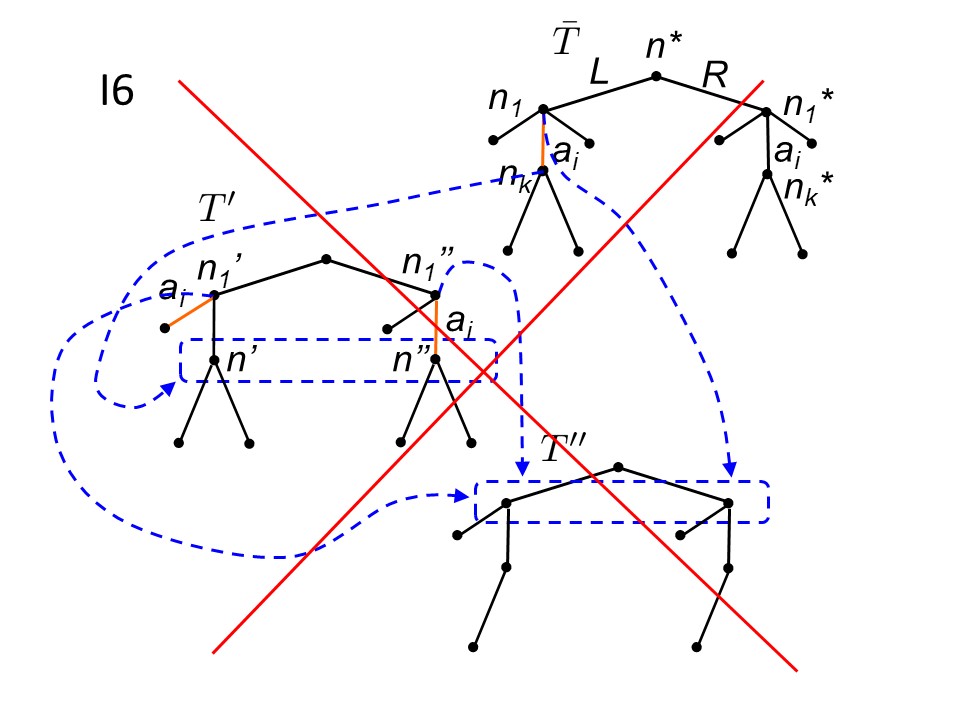} \includegraphics[scale=.3]{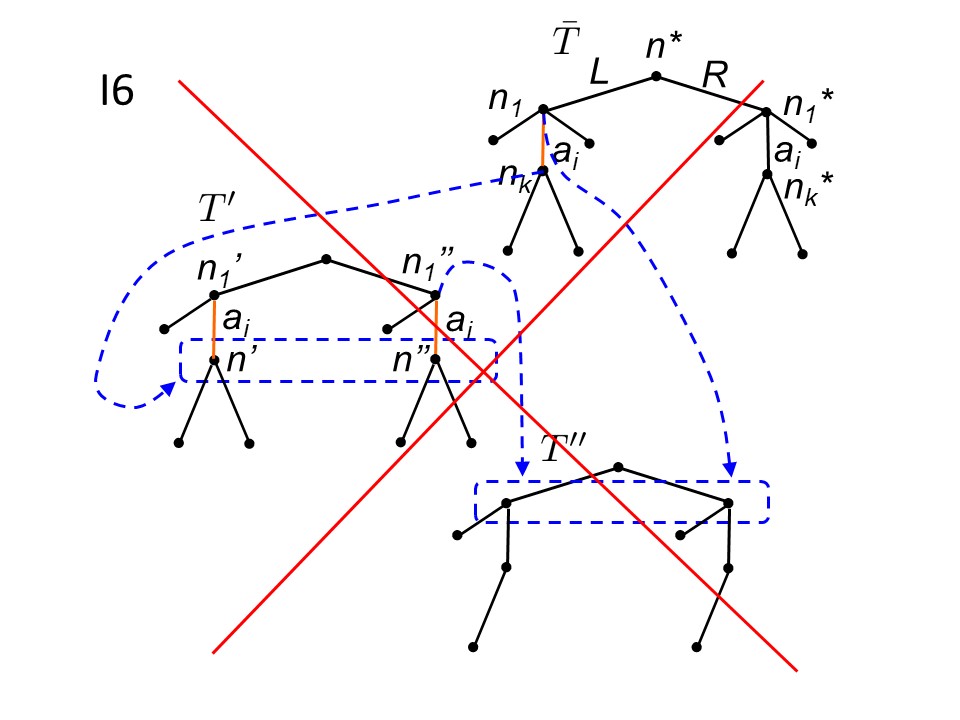}
\end{center}
\end{figure}

This property is illustrated with an example in Figure~\ref{I6_1} and two counterexamples in Figure~\ref{I6_2}. Suppose the tree modeling all physical moves is given by $\bar{T}$, and $T'$, $T''$ are subtrees with $T' \preceq T'' \preceq \bar{T}$. To avoid clutter, we have included only the information sets of interest in the diagram which are drawn with blue dashed lines. In each tree, player 1 is the only player active at the initial decision node and player 2 is the only player active at all other decision nodes, for example with respect to $\bar{T}$ player 1 is active at node $n^*$ and player 2 is active at nodes $n'_1 , n''_1, n'$, and $n''$. The other nodes are terminal nodes. We restrict attention to the path where player 1 plays $L$ at node $n^*$ and player 2 plays $a_i$ at $n_1$, which is the action that has been colored orange. The information set at $n_k$ contains two nodes, $n'$ and $n''$ both of which are in $T'$. Perfect recall requires 1) that both $n'$ and $n''$ have a decision node of player 2 that precedes them in $T'$ where $a_i$ is played along the path, a condition which is satisfied by $n'_1$ and $n''_1$ respectively; and 2) that these nodes $n'_1$ and $n''_1$ have the same information set as $n_1$, which is satisfied as all three nodes have the same information set in $T''$. Perfect recall requires these two conditions to be met for any two nodes of a player whenever one precedes another in a tree. The diagrams in Figure~\ref{I6_2} are two examples in which the condition of perfect recall is not satisfied. In the left diagram, $n'$ does not have a decision node of player 2 that precedes it in $T'$ where $a_i$ is played. In the right diagram, the node $n'_1$ does not have the same information set as $n_1$. 

It is known that in standard games of the extensive form without unawareness, perfect recall is necessary for the ``playability'' of strategies\footnote{A mixed strategy may require for instance to mix between strategy ``play left after having played left'' and ``play right after having played right''. Without perfect recall, such a strategy is not playable because when the player moves the second time, she has forgotten whether she played right or left before.} and the existence of Nash equilibrium in behavior strategies; see for instance Wichardt (2008) (see also an earlier example by Luce and Raiffa, 1957, p. 160). Since standard games of the extensive form are special cases of games of the extensive form with unawareness, such arguments apply also to games of the extensive form with unawareness.


In what follows, we provide two alternate characterizations of the perfect recall condition. They lend support to our notion of perfect recall and could also provide additional insight as to how it operates. The first characterization comes in the form of a useful remark relating our definition to the one provided in Selten (1975). For two nodes $n,n'$ where $T_n = T_{n'}$, define $n \lessdot_a n'$ to hold if action $a$ is taken at $n$ in order to get to $n'$.

\begin{rem}\label{records1} A game of the extensive form with unawareness satisfies perfect recall (I6) if and only if for any player $i \in I$, $n_1, n_k, n'_k \in \mathbf{D}_i$ such that $n'_k \in h_i (n_k)$, we have that $n_1 \lessdot_{a_i} n_k$ for some $a_i \in A_i (n_1)$ implies $n'_1 \lessdot_{a_i} n'_{k}$ for some $n'_1$ and $h_i (n'_1) = h_i (n_1)$.
\end{rem}

\noindent \textsc{Proof. } ``$\Rightarrow$'': Suppose by contradiction that I6 holds, and for some player $i \in I$, $n_1, n_k, n'_k \in \mathbf{D}_i$ and $n_1 \lessdot_{a_i} n_k$ for some ${a_i} \in A_i (n_1)$ and $n'_k \in h_i (n_k)$, but there does not exist $n'_1$ such that $n'_1 \lessdot_{a_i} n'_{k}$ with $h_i (n'_{1}) = h_i (n_1)$. By I6 for each $n''_{\ell} \in h_i (n_k) \subseteq T'_{h_i (n_k)}$ there exists a path $n''_1, n''_2, ..., n''_{\ell}$ such that $h_i(n_1) = h_i(n''_{1})$ and the action taken at $n''_1$ along the path $n''_1 , n_2''$ is $a_i$ (i.e., $n''_1 \lessdot_{a_i} n'_{k}$), a contradiction.

``$\Leftarrow$'':  Suppose not. Then there are two nodes $n_1$ and $n_k$ with $n_1 \neq n_k$ with the path $n_1, n_2, ..., n_k$ such that at $n_1$ player $i$ takes action $a_i$ along the path, but there is some $n' \in h_i(n_k)$ where there is no node $n_1' \neq n'$ where there is a path $n_1', n_2', ..., n_{\ell}' = n'$, $h_i(n_1') = h_i(n_1)$ and player $i$ takes action $a_i$ at $n_1'$ along the path. But this means that for player $i$, $n'_{\ell} \in h_i (n_k) $ and $n_1 \lessdot_{a_i} n_k$ but there is no $n'_1$ such that $n'_1 \lessdot_{a_i} n'_{\ell}$ with $h_i (n'_1) = h_i (n_1)$, a contradiction. \hfill $\Box$\\

This is exactly the definition provided in Selten (1975), except now $T_{n_k}$ and $T_{h_i (n_k)}$ may not be the same tree as we consider a forest of trees rather than just one tree. That is, with respect to our definition of perfect recall I6, the nodes $n_1$ and $n_k$ are in the same tree, the nodes $n'_1$ and $n'_{\ell}$ are in the same tree as the information set $h_i (n_k)$ (equivalently $h_i (n'_\ell)$), however these two may not be the same tree. We occasionally utilize this version of perfect recall to streamline proofs.

Perfect recall may also be understood as the requirement that players do not forget their experience over the course of the game. This can be made explicit. For any player $i \in I$ and decision node of that player $n \in \mathbf{D}_i$, let $E_i(n)$ denote the record of player $i$'s experience along the path to $n$ (not including $h_i(n)$). I.e., $E_i(n)$ is the sequence of pairs $(h_i, a_i)$ of player $i$'s information sets and the action taken at these information sets in order of how they are encountered along the path to $n$. Perfect recall is now characterized as follows:

\begin{rem}\label{records} A game of the extensive form with unawareness satisfies perfect recall (I6) if and only if for any player $i \in I$, $n \in \mathbf{D}_i$, $n' \in h_i(n)$ implies $E_i(n') = E_i(n)$.
\end{rem}

\noindent \textsc{Proof. } ``$\Rightarrow$'': Consider the non-trivial case $n' \neq n$. Suppose by contradiction that $E_i(n') \neq E_i(n)$. By U0 there are two possible scenarios. The first is when the nodes $n,n'$ are in the same tree, i.e., $n, n' \in h_i(n) \subseteq T'$, whereupon we obtain an immediate contradiction with I6. The second is when the nodes $n,n'$ are in different trees, i.e., $n \in T$, $n' \in h_i(n) \subseteq T'$ where $T \neq T'$, $T' \preceq T$. First it follows by U1 there is $n_{T'} \in h_i(n) \subseteq T'$. Then by I6, for any node $n_1 \in T$, any action $a_i \in A_i  (n_1)$ such that $n_1 \lessdot_{a_i} n$, there is $(n_{1})_{T'} \in T'$ such that $(n_1)_{T'} \lessdot_{a_i} n_{T'}$ and $h_i (n_1) = h_i ((n_1)_{T'})$. Since $T' \preceq T$ we also have that for any node $(n_1)_{T'} \in T'$, any action $a_i \in A_i ((n_1)_{T'})$ such that $(n_1)_{T'} \lessdot_{a_i} n_{T'}$, there is $n_{1}$ such that $n_1 \lessdot_{a_i} n$, and by I3 and I5 it must be that $h_i ((n_1)_{T'}) = h_i (n_1)$. Hence we have a contradiction.

``$\Leftarrow$'':  Suppose not. Then there are two nodes $n_1$ and $n$ with $n_1 \neq n$ and action $a_i \in A_i (n_1)$ such that $n_1 \lessdot_{a_i} n$, but there is some $n' \in h_i(n)$ where there is no node $n_1' \neq n'$ such that $n'_1 \lessdot_{a_i} n'$. But this just means that player's records of experience are different in nodes $n$ and $n'$, i.e., $E_i(n') \neq E_i(n)$, a contradiction. \hfill $\Box$\\

This characterization in terms of `experience' in Remark~\ref{records} helps by being another familiar way of understanding perfect recall, and is also useful in helping us prove the main theorem of the paper. We elaborate more on perfect recall in Section~\ref{KT_section}.

We denote by $H_{i}$ the set of player $i$'s information sets in all trees. For an information set $h_{i} \in H_{i}$, we denote by $T_{h_{i}}$ the tree containing $h_{i}$. For two information sets $h_{i}, h_{i}^{\prime}$ in a given tree $T,$ we say that $h_{i}$ precedes $h_{i}^{\prime }$ (or that $h_{i}^{\prime }$ succeeds $h_{i}$) if for every $n^{\prime} \in h_{i}^{\prime }$ there is a path $n,...,n^{\prime }$ in $T$ such that $n \in h_{i}$. We denote it by $h_{i}\rightsquigarrow h_{i}^{\prime }$.

The following property is implied by I2 and I4 (see Heifetz, Meier, and Schipper, 2013, Remark 1): If $n^{\prime}, n^{\prime \prime } \in h_{i}$ where $h_{i}=h_{i}\left( n\right)$ is an information set for some $n \in \mathbf{D}_i$, then $A_i(n') = A_i(n'')$. Hence, if $n \in h_i$ we write also $A_i(h_i)$ for $A_i(n)$.

Properties U0, U1, I2, and I6 imply no absent-mindedness. This follows directly from Heifetz, Meier, and Schipper (2013, Remark 2).
\begin{itemize}
\item[] No Absent-mindedness: No information set $h_{i}$ contains two distinct nodes $n, n'$ on the same path in some tree.
\end{itemize}

The Perfect recall property I6 therefore guarantees that with the precedence relation $\rightsquigarrow $ player $i$'s information sets $H_{i}$ form an unrooted tree: For every information set $h_{i}^{\prime }\in H_{i}$, the information sets preceding it $\left\{ h_{i} \in H_{i}: h_{i} \rightsquigarrow h_{i}^{\prime }\right\}$ are totally ordered by $\rightsquigarrow$.

Confined awareness (U0) and Perfect recall (I6) imply that a player cannot become unaware during the play (see Heifetz, Meier, and Schipper, 2013, Remark 6).
\begin{itemize}
\item[DA] Awareness may only increase along a path: If there is a path $n, \dots, n^{\prime}$ in some subtree $T''$ such that player $i$ is active in $n$ and $n^{\prime }$, and $h _{i}\left( n\right) \subseteq T$ while $h_{i}\left( n^{\prime }\right) \subseteq T^{\prime }$, then $T \preceq T^{\prime}$.
\end{itemize}

To model unawareness proper, we impose as in Heifetz, Meier, and Schipper (2013) additional properties. They parallel properties of static unawareness structures in Heifetz, Meier, and Schipper (2006):
\begin{itemize}
\item[U4] Subtrees preserve ignorance: If $T \preceq T' \preceq T''$, $n \in T''$, $h_{i}(n) \subseteq T$ and $T'$ contains the copy $n_{T'}$ of $n$, then $h_{i}(n_{T'}) = h_{i}( n)$.

\item[U5] Subtrees preserve knowledge: If $T \preceq T' \preceq T''$, $n \in T''$, $h_{i}(n) \subseteq T'$ and $T$ contains the copy $n_{T}$ of $n$, then $h_{i}(n_{T})$ consists of the copies that exist in $T$ of the nodes of $h_{i}(n)$.
\end{itemize}
It is known that U5 implies U3, see Heifetz, Meier, and Schipper (2013, Remark 3):
\begin{itemize}
\item[U3] Subtrees preserve awareness: If $n \in T'$, $n \in h_i(n)$, $T \preceq T'$, and $T$ contains a copy $n_T$ of $n$, then $n_T \in h_i(n_T)$.
\end{itemize}

\begin{figure}[h!]
\caption{Properties U3 to U5}
\label{propertiesU}
\begin{center}
\includegraphics[scale=0.25]{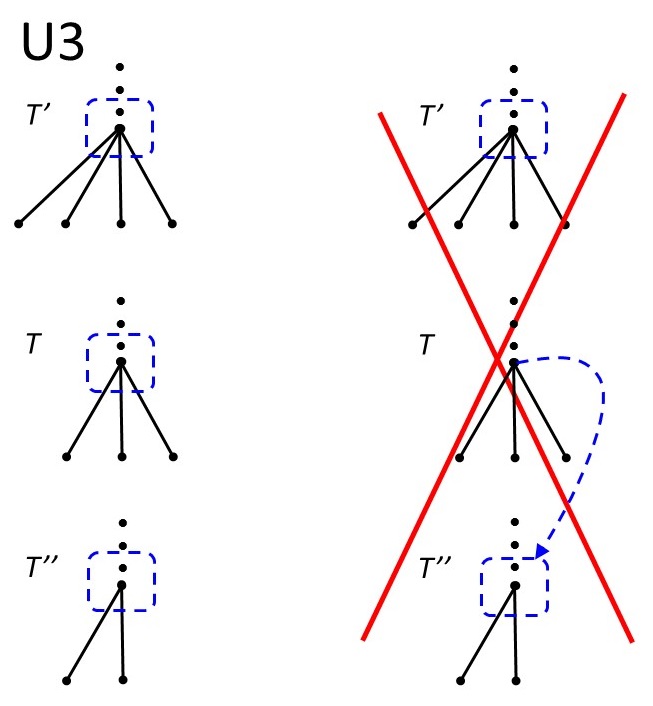}
\bigskip

\bigskip

\includegraphics[scale=0.25]{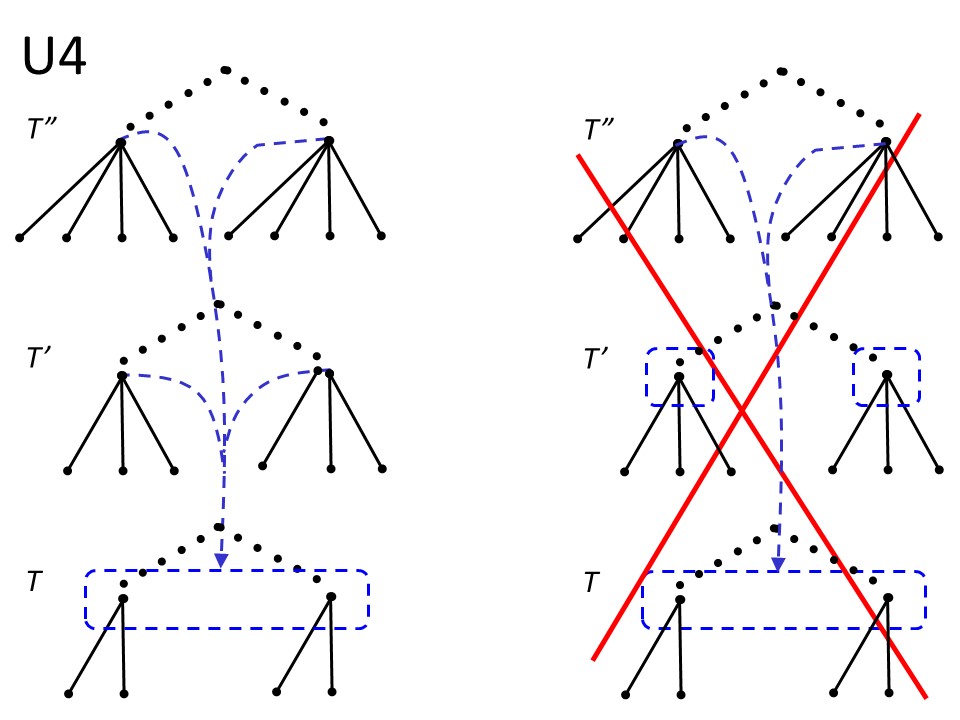} \qquad \qquad \includegraphics[scale=0.25]{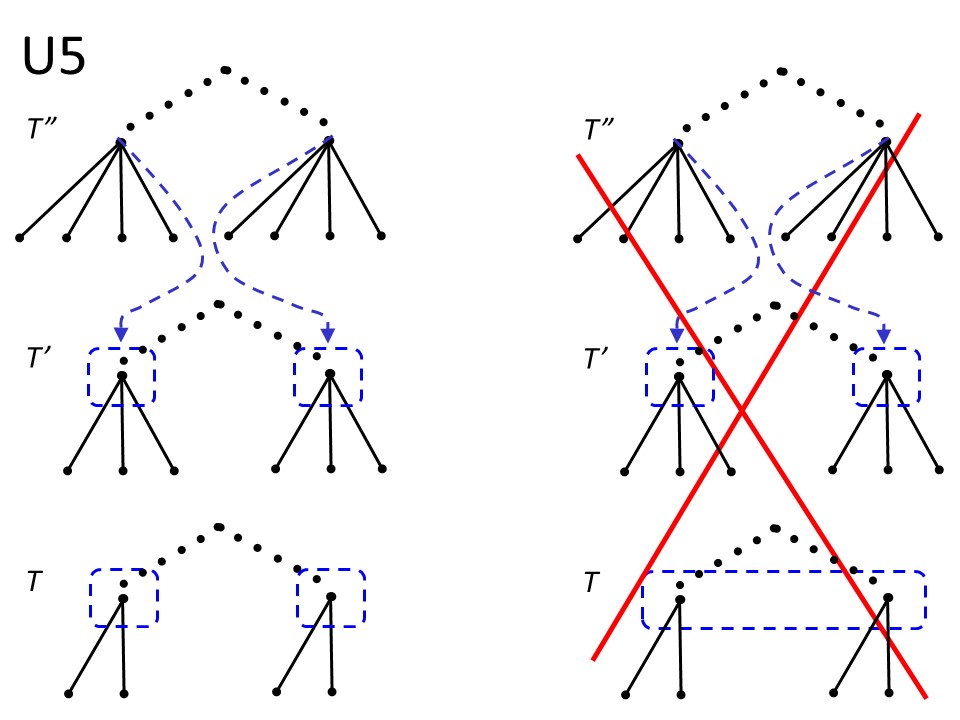}
\end{center}
\end{figure}

Properties U3 to U5 are illustrated in Figure~\ref{propertiesU} with an example and counterexample each.

For trees $T, T^{\prime } \in \mathbf{T}$ we denote by $T \rightarrowtail T^{\prime }$ whenever for some node $n\in T$ and some player $i\in P(n)$ it is the case that $h_{i}(n) \subseteq T'$. Denote by $\hookrightarrow $\ the transitive closure of $\rightarrowtail$. That is, $T \hookrightarrow T^{\prime \prime}$ if and only if there is a sequence of trees $T, T^{\prime }, \dots, T^{\prime \prime } \in \mathbf{T}$ satisfying $T \rightarrowtail T^{\prime} \rightarrowtail \dots \rightarrowtail T^{\prime\prime}$.

A \emph{game of the extensive form with unawareness} $\Gamma $ consists of a join-semilattice $\mathbf{T}$ of subtrees of a tree $\bar{T}$ satisfying properties 1--3 above, along with information sets $h_i(n)$ for every $n \in T$ with $T \in \mathbf{T}$ and $i \in P(n)$, and payoffs satisfying properties U0, U1, U4, U5, and I2-I5 above. We say that the game satisfies perfect recall if I6 also holds for all players $i \in I$.

For any game of the extensive form with unawareness $\Gamma$ with set of trees $\mathbf{T}$, for any tree $T \in \mathbf{T}$, the $T$\emph{-partial game} is the join-semisublattice of trees including $T$ and also all trees $T^{\prime}$ in $\Gamma$ satisfying $T \hookrightarrow T^{\prime}$, with information sets as defined in $\Gamma$. A $T$-partial game is a game of the extensive form with unawareness, i.e., it satisfies all properties 0--3, U0, U1, U4, U5, and I2-I5 above, and we say that the game satisfies perfect recall if I6 also holds for all players $i \in I$.

We denote by $H_{i}^{T}$ the set of player $i$'s information sets in the $T$-partial game, $T \in \mathbf{T}$. This set contains not only $i$'s information sets in the tree $T$ but also in all trees $T' \in \mathbf{T}$ with $T \hookrightarrow T'$.

\section{Strategies\label{strategies}}

For any collection of sets $(X_i)_{i \in I^0}$ we denote by
$$X := \prod_{i \in I^0} X_{i}, \quad X_{-i} := \prod_{j \in I^0 \setminus \{i\}} X_{j}$$ with typical elements $x$ and $x_{-i}$ respectively. For any collection of sets $(X_i)_{i \in I^0}$ and any tree $T \in \mathbf{T}$, we denote by $X_i^T$ the set of objects in $X_i$ restricted to the tree $T$ and analogously for $X^T$ and $X^T_{-i}$, where ``restricted to the tree $T$'' will become clear from the definitions below.

A \emph{pure strategy} for player $i \in I$,
\begin{equation*}
s_{i} \in S_{i} := \prod_{h_{i}\in H_{i}} A(h_{i})
\end{equation*}
specifies an action of player $i$ at each of her information sets $h_{i}\in H_{i}$. We let
\begin{equation*}
s_0 \in S_0 := \prod_{n \in \mathbf{D}_0} A_0(n)
\end{equation*} denote the ``strategy'' of nature, with $\mathbf{D}_0$ denoting the ``decision'' nodes of nature.

\begin{figure}[h!]
\begin{center}
\caption{Action induced by a strategy\label{strategy}}
\includegraphics[scale=0.4]{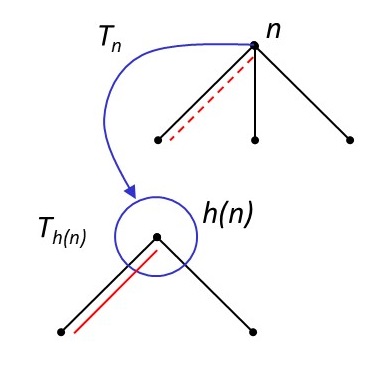}
\end{center}
\end{figure}

With the strategy $s_{i}$, at node $n \in D_{i}^{T_n}$ define player $i$'s action at $n$ to be $s_i(h_{i}(n))$, for $i \in I$. Thus, by U1 and I4 the strategy $s_{i}$ specifies what player $i \in I$ does at each of her active nodes $n \in D_{i}^{T_n}$, both in the case that $n \in h_i(n)$ \emph{and} in the case that $h_i(n) $ is a subset of nodes of a tree which is distinct from the tree $T_{n}$ to which $n$ belongs. In the first case, when $n \in h_i(n)$, we can interpret $s_i(h_i(n))$ as the action chosen by player $i$ in node $n$. In the second case, when $n \notin h_i(n)$, $s_i(h_i(n))$ cannot be interpreted as the action chosen ``consciously'' by player $i$ in $n$ since she is not even aware of $T_n$. Instead, her state of mind at $n$ is given by her information set $h_i(n)$ in a tree lower than $T_n$ (denoted by $T_{h_i(n)}$). Thus, $s_i(h_i(n))$ is the physical move of player $i$ in $n$ in tree $T_n$ induced by her ``consciously'' chosen action at her information set $h_i(n)$ in tree $T_{h_i(n)}$ (with $T_n \succ T_{h_i(n)}$). As an example, consider the game in Figure~\ref{strategy}. The information set at node $n$ in tree $T_n$ lies in the lower tree $T_{h(n)}$ that misses the action ``middle''. This is indicated by the blue arrow and disk. When the player chooses ``left'' in $T_{h(n)}$ (as indicated by the red solid line beside the left edge), it induces also an action ``left'' at node $n$ in tree $T_n$ (as indicated by the red dashed line beside the left edge).

In a game of the extensive form with unawareness $\Gamma$ the tree $\bar{T} \in \mathbf{T}$ represents the physical paths in the game; every tree in $\mathbf{T}$ that contains an information set represents the subjective view of the feasible paths in the mind of a player, or the view of the feasible paths that a player believes that another player may have in mind, etc. Moreover, as the actual play in $\bar{T}$ unfolds, a player may become aware of paths of which she was unaware earlier, and the way she views the game may be altered. Thus, in a game of the extensive form with unawareness, a strategy cannot be conceived as an ex ante plan of action. Formally, a strategy of player $i$ is a list of answers to the questions ``what would player $i \in I$ do if $h_{i}$ were the set of nodes she considered as possible?'', for $h_{i}\in H_{i}$ (and analogous for nature). A strategy of a player becomes meaningful as an object of beliefs of other players. How ``much'' of a player's strategy other players can conceive depends on their awareness given by the tree in which their information set is located. This leads to the notion of $T$-partial strategy. For a strategy $s_{i} \in S_{i}$ and a tree $T \in \mathbf{T}$, we denote by $s_{i}^{T}$ the strategy in the $T$-partial game induced by $s_{i}$ (i.e., $s_{i}^{T}\left( h_{i}\right) = s_{i}\left(h_{i}\right)$ for every information set $h_{i} \in H_i^T$ of player $i$ in the $T$-partial game). (Recall that $H_i^T$ not only contains information sets in the tree $T$ but also in trees $T' \in \mathbf{T}$ with $T' \preceq T$.)

A \emph{mixed strategy} of player $i \in I^0$, $\sigma_{i} \in \Delta(S_{i})$, specifies a probability distribution over player $i$'s set of pure strategies. With this notation, we let $\sigma_0$ be the probability distribution over ``strategies'' of nature. As mentioned already in the introduction, we don't consider mixed (or behavior) strategies necessarily as an object of choice of players but rather a conjecture over how a player would play.

A \emph{behavior strategy} for player $i \in I$,
\begin{equation*} \beta_i \in B_i := \prod_{h_i \in H_i} \Delta(A_i(h_i))
\end{equation*}
is a collection of independent probability distributions, one for each of player $i$'s information sets $h_{i}\in H_{i}$, where $\beta_i(h_i)$ specifies a mixed action in $\Delta(A_{h_i})$. With the behavior strategy $\beta_i$, at node $n \in D_i^{T_n}$ define player $i$'s mixed action at $n$ to be $\beta_i(h_i(n))$. Thus, the behavior strategy $\beta_i$ specifies the mixed action of player $i \in I$ at each of her active decision nodes $n \in D_i^{T_n}$, both in the case that $n \in h_i(n)$ and in the case that $h_i(n)$ is a subset of nodes of a tree which is distinct from the tree $T_n$ to which $n$ belongs. It may be the case that $A_i(n) \supset A_i(h_i(n))$. Yet, we have automatically that $\beta_i$ does not assign probabilities to actions in $A_n \setminus A_{h_i(n)}$. (I.e., at the decision node $n$ of the richer tree $T_n$ player $i$ may have more actions than she is aware of at $h_i(n)$. In such a case, she is unable to use actions that she is unaware of.) With respect to nature, we let $\beta_0 \in B_0 = \prod_{n \in D_0} \Delta(A_0(n))$.

We say that a strategy profile $s=\left( s_{j}\right) _{j\in I}\in S$ \emph{reaches a node} $n\in T$ if $n$ is on the path of play in $T$ given the players' actions and nature's moves $\left(s_{j}^{T}\left( h_{j}\left( n^{\prime }\right) \right)\right)_{j \in P(n')}$ in nodes $n^{\prime }\in T$. Notice that by property (I4) (``no imaginary actions''), $s_{j}^{T}\left( h _{j}\left( n^{\prime }\right) \right) _{j\in I}$ is indeed well defined: even if $h_{j}\left(n'\right) \nsubseteq T$ for some $n^{\prime }\in T$, $\left(s_{j}^{T}\left( h_{j}\left( n'\right) \right)\right)_{j \in P(n')}$ is a profile of actions which is actually available in $T$ to the active players $j \in P(n')$ and possibly nature at $n'$. We say that a strategy profile $s\in S$ \emph{reaches} the information set $h_{i}\in H_{i}$ if $s$ reaches some node $n\in h_{i}$. We say that the strategy $s_{i}\in S_{i}$ \emph{allows the node $n$ to be reached} if there is a strategy profile $s_{-i}\in S_{-i}$ of the other players (and possibly nature) such that the strategy profile $\left( s_{i},s_{-i}\right)$ reaches $n$. Analogously, we say that the strategy profile $s_{-i} \in S_{-i}$ \emph{allows the information set $h_{i}$ to be reached} if there exists a strategy $s_{i}\in S_{i}$ such that the strategy profile $\left( s_{i},s_{-i}\right)$ reaches $h_{i}$. For each player $i \in I$, denote by $H_i(s)$ the set of information sets of $i$ that are reached by the strategy profile $s$. This set may contain information sets in more than one tree.

We extend the definitions of information set reached to mixed and behavior strategies in the obvious way by considering nodes/information sets reached with strict positive probability.

\section{Kuhn's Theorem\label{KT_section}}

In games of the extensive form without unawareness but with perfect recall, Kuhn's Theorem asserts that for every mixed strategy profile there is an equivalent behavior strategy profile. Kuhn's Theorem can be extended to games of the extensive form with unawareness using a notion of equivalence based on the notion of reaching nodes. For any node $n$, any player $i \in I^0$, and any opponents' profile of strategies $s_{-i}$ (including nature if any), let $\rho(n \mid \beta_i, s_{-i})$ and $\rho(n \mid \sigma_i, s_{-i})$ denote the probability that $(\beta_i, s_{-i})$ and $(\sigma_i, s_{-i})$ reach node $n$, respectively. For any player $i \in I^0$, a mixed strategy $\sigma_i$ and a behavior strategy $\beta_i$ are \emph{equivalent} if for every profile of opponents' strategies $s_{-i} \in S_{-i}$ and every node $n \in \mathbf{N}$ of the game of the extensive form with unawareness $\rho(n \mid \sigma_i, s_{-i}) = \rho(n \mid \beta_i, s_{-i})$. To simplify the exposition, we also say that mixed and behavior strategies are \emph{equivalent}, or, that an \emph{equivalence} between mixed and behavior strategies is obtained, when in the game for every mixed strategy there is an equivalent behavior strategy and vice versa.

Let $S_i(n)$ be the set of all strategies of player $i$ that allows $n$ to be reached. That is, if $n \in T$ then $s_i \in S_i(n)$ if and only if there exist $s_{-i} \in S_{-i}$ such that the profile $(s_j^T(h_j(n')))_{j \in P(n')}$ prescribed for $n'\lessdot n$ reaches $n$.

In the following, we extend for better comparison a textbook proof of Kuhn's Theorem (e.g., Maschler, Solan, and Zamir, 2013, Chapter 6) to the more complicated set up of games of the extensive form with unawareness. It allows us to show which modifications are necessary as compared to standard games, and emphasizes that the basic idea of the proof remains the same although the setting is considerably more complicated. We start with a lemma that is crucial for the proof of the main theorem.

\begin{lem}\label{Kuhnlemma} Consider a game of the extensive form with unawareness $\Gamma$. If $\Gamma$ satisfies perfect recall (i.e., I6), then for any player $i \in I$, $n, n' \in N$ such that $n' \in h_i(n) \in H_i$, we have $S_i(n) = S_i(n')$.
\end{lem}

Note the difference to standard games. First, games of the extensive form with unawareness involve forests of trees rather than just trees. Second, the perfect recall property applies now to information sets across trees. Third, the sets $S_i(n)$ are different from corresponding sets in standard games because $n$ may be in a tree different from the tree ``housing'' the information set $h_i(n)$ of player $i$ at $n$. This is significant because strategies are defined for each information set (rather than decision node) of player $i$.\\

\noindent \textsc{Proof of Lemma~\ref{Kuhnlemma}. } Since $\Gamma$ satisfies perfect recall, we have by Remark~\ref{records} that $n' \in h_i(n)$ implies $E_i(n') = E_i(n)$. Thus, the same information sets of player $i$ that are reached along the path to $n$ are reached along the path to $n'$ (even though the information sets might appear in a subtree lower than the one containing the path to $n$). Moreover, at each of those information sets the same action is required to move along the path to $n$ as to move along to the path to $n'$. Hence, any strategy that allows $n$ to be reached also allows $n'$ to be reached and vice versa. \hfill $\Box$\\

We are now ready to state an extension of Kuhn's Theorem to games of the extensive form with unawareness.

\begin{theo}\label{Kuhn} In every game of the extensive form with unawareness, if player $i$ has perfect recall, then for every mixed strategy of player $i$ there exists an equivalent behavior strategy.
\end{theo}

\noindent \textsc{Proof. } The theorem is proved in three steps. The first step defines the candidate of the behavior strategy. The second step shows that it is well-defined. The third step shows it to be equivalent to the mixed strategy.

Let $\sigma_i$ be a mixed strategy of player $i$.

First, we define a candidate for the equivalent behavior strategy. Let $n \in \mathbf{D}_i$. Since $n \in \mathbf{D}_i$ we have $A_i(n) \neq \emptyset$. For any action $a_i \in A_i(n)$ of player $i$ at $n$, define $S_i(n, a_i) := \{s_i \in S_i(n) : s_i(h_i(n)) = a_i\}$. That is, any strategy in $S_i(n, a_i)$ allows $n$ to be reached and also prescribes action $a_i$ at information set $h_i(n)$. This definition makes sense: First, recall that strategies ascribe actions to information sets (rather than nodes). Moreover, by definition of $\Gamma$ there is an information set of player $i$ at $n$ that we denote by $h_i(n)$. There are two cases: First, $n \in h_i(n)$. In this case, for any $n' \in h_i(n)$ we have $A_i(n) = A_i(n')$ (Heifetz, Meier, and Schipper, 2013, Remark 1). Thus, we write $A_i(h_i(n))$ for actions available at any node in $h_i(n)$. Second, $n \notin h_i(n)$. (That's the case when $n$ is in a tree more expressive than $h_i(n)$.) By I4 (No imaginary actions), for any $n' \in h_i(n)$ we have $A_i(n') \subseteq A_i(n)$. Note also that $A_i(n') \neq \emptyset$ since $n' \in \mathbf{D}_i$. If $a_i \in A_i(n) \setminus A_i(n')$, then $S_i(n, a_i) = \emptyset$ since no strategy of player $i$ can ascribe an action to $n$ that is not available at $h_i(n)$. Hence, in the following we consider sets $S_i(n, a_i)$ for $a_i \in A_i(h_i(n))$.

If player $i$'s mixed strategy assigns strict positive probability to strategies allowing $n$ to be reached, i.e., if $\sum_{s_i \in S_i(n)} \sigma_i(s_i) > 0$, then define for each action $a_i \in A_i(h_i(n))$,
\begin{eqnarray}\label{star} \beta_i(h_i(n))(a_i) := \frac{\sum_{s_i \in S_i(n, a_i)} \sigma_i(s_i)}{\sum_{s_i \in S_i(n)} \sigma_i(s_i)}.
\end{eqnarray} Otherwise, if $\sum_{s_i \in S_i(n)} \sigma_i(s_i) = 0$, define $\beta_i(h_i(n))$ in an arbitrary way provided that it constitutes a probability measure over actions available at $h_i(n)$. E.g., for all $a_i \in A_i(h_i(n))$,
\begin{eqnarray}\label{starstar} \beta_i(h_i(n))(a_i) := \frac{1}{| A_i(h_i(n)) |}.
\end{eqnarray}

Second, we want to show that $\beta_i$ is well-defined. I.e., for each of player $i$'s information sets $h_i \in H_i$, $\beta_i(h_i)$ is a probability measure on $A_i(h_i)$. Moreover, $\beta_i$ is independent of player $i$'s decision nodes. For the latter, it suffices to demonstrate it for the case $\sum_{s_i \in S_i(n)} \sigma_i(s_i) > 0$. Since $\Gamma$ has perfect recall, i.e., $\Gamma$ satisfies I6, we have by Lemma~\ref{Kuhnlemma} that for any $n' \in h_i(n)$, $S_i(n') = S_i(n)$. Again, since $n' \in h_i(n)$, $S_i(n, a_i) = S_i(n', a_i)$ for all $a_i \in A_i(h_i)$. Observe that both the numerator and the denominator of the right-hand side of Equation~(\ref{star}) are independent of nodes in $h_i(n)$.

To show that for all $n \in \mathbf{D}_i$, $\beta_i(h_i(n))$ is a probability measure over $A_i(h_i(n))$, note first that, in the case of $\sum_{s_i \in S_i(n)} \sigma_i(s_i) = 0$, this follows directly from Equation~(\ref{starstar}).

If $\sum_{s_i \in S_i(n)} \sigma_i(s_i) > 0$, then Equation~(\ref{star}) defines a probability distribution over $A_i(h_i)$. To see this note that since $\sigma_i(s_i) \geq 0$ for all $s_i \in S_i$, both the numerator and the denominator are non-negative and hence $\beta_i(h_i)(a_i) \geq 0$ for all $a_i \in A_i(h_i)$. For any $n \in \mathbf{D}_i$ and $a_i, a_i' \in A_i(h_i(n))$ with $a_i \neq a_i'$, $S_i(n, a_i) \cap S_i(n, a_i') = \emptyset$. Moreover, $\bigcup_{a_i \in A_i(h_i(n))} S_i(n, a_i) = S_i(n)$. Thus, $\sum_{a_i \in A_i(h_i(n))} \sum_{s_i \in S_i(n, a_i)} \sigma_i(s_i) = \sum_{s_i \in S_i(n)} \sigma_i(s_i)$. It follows that $\sum_{a_i \in A_i(h_i)} \beta_i(h_i)(a_i) = 1$.

The third and last step is to show that the behavior strategy $\beta_i$ is equivalent to the mixed strategy $\sigma_i$. The probability $\rho_i$ that node $n$ is allowed to be reached given mixed strategy $\sigma_i$ of player $i$ is: $\rho_i (n \mid \sigma_i) := \Sigma_{s_i \in S_i (n)} \sigma_i (s_i)$. When $S_i (n) = \emptyset$, $\rho_i (n \mid \sigma_i) := 0$. The probability $\rho_{-i}$ that node $n$ is allowed to be reached given some profile of player $i$'s opponents' (pure) strategies $s_{-i} \in S_{-i}$, $\rho_{-i} (n \mid s_{-i})$, equals to $1$ if $s_{-i} \in S_{-i} (n)$ and $0$ if $s_{-i} \notin S_{-i} (n)$. Then, as the strategies between players are independent, we obtain that $\rho(n \mid \sigma_i, s_{-i}) = \rho_i (n \mid \sigma_i) \times \rho_{-i} (n \mid s_{-i}) = \Sigma_{s_i \in S_i (n)} \sigma_i (s_i)$ if $s_{-i} \in S_{-i} (n)$ and $0$ otherwise. We obtain an analogous result for $\beta_i$ due to the independence of strategies between players. For any $n \in \mathbf{N}$ we also have $\rho(n \mid \beta_i, s_{-i}) = \rho_{i} (n \mid \beta_i)$ if $s_{-i} \in S_{-i} (n)$ and $0$ otherwise. As a result, in what follows we can focus without loss of generality on the case where $s_{-i} \in S_{-i} (n)$.

Fix a node $n \in \mathbf{D}_i \cup \mathbf{Z}$ and let $n_i^1, n_i^2, ..., n_i^L$ be a sequence of decision nodes of player $i$ along the path from the root to $n$, not including $n$. By definition of $\Gamma$, there exists an information set of player $i$ for each of the decision nodes $n_i^1, n_i^2, ..., n_i^L$. If $L = 0$, then player $i$ has no information sets on the path from the root to $n$ (not including $n$). In such a case, $S_i(n) = S_i$ and for any $s_{-i} \in S_{-i}$ we naturally define $\rho_i(n \mid \beta_i) = 1$. We therefore have $\rho(n \mid \beta_i, s_{-i}) = 1$ for $s_{-i} \in S_{-i} (n)$. Also, in this case $\rho(n \mid \sigma_i, s_{-i}) = \rho_i (n \mid \sigma_i) = \sum_{s_i \in S_i(n)} \sigma_i(s_i) = \Sigma_{s_i \in S_i} \sigma_i(s_i) = 1$ for $s_{-i} \in S_{-i} (n)$. Hence, $\beta_i$ and $\sigma_i$ are equivalent in this case.

Suppose now the case $L > 0$. Let $a_i^{n^{\ell}_i} \in A_i(h_i(n^{\ell}_i))$ denote the action of player $i$ at node $n^{\ell}_i$, $\ell = 1, ..., L$, that leads to $n_i^{\ell + 1}$ in the case $\ell = 1, ..., L-1$ and to $n$ in the case $\ell = L$. Again, assume without loss of generality that $s_{-i} \in S_{-i} (n)$.

We have
\begin{eqnarray} \rho(n \mid \beta_i, s_{-i}) = \rho_i (n \mid \beta_i)  =  \prod_{\ell = 1}^{L} \beta_i(h_i(n^{\ell}_i))(a_i^{n^{\ell}_i}).
\end{eqnarray} Assume without loss of generality that $\sigma_i$ allows $n$ to be reached (otherwise $\rho_i (n \mid \sigma_i) = \rho_i (n \mid \beta_i) = 0$). By definition of $\beta_i$,
\begin{eqnarray} \rho(n \mid \beta_i, s_{-i}) & = & \prod_{\ell = 1}^{L} \frac{\sum_{s_i \in S_i(n^{\ell}_i, a_i^{n^{\ell}_i})} \sigma_i(s_i)}{\sum_{s_i \in S_i(n^{\ell}_i)} \sigma_i(s_i)},
\end{eqnarray} which is well-defined since $\sigma_i$ is assumed to allow $n$ to be reached and therefore also allows $n_i^1, ..., n_i^L$ to be reached.

Note that $S_i(n_i^{\ell + 1}) = S_i(n_i^{\ell}, a_i^{\ell})$. Thus
\begin{eqnarray*} \sum_{s_i \in S_i(n_i^{\ell}, a_i^{\ell})} \sigma_i(s_i) & = & \sum_{s_i \in S_i(n^{\ell + 1}_i)} \sigma_i(s_i)
\end{eqnarray*} and
\begin{eqnarray} \rho(n \mid \beta_i, s_{-i}) & = & \prod_{\ell = 1}^{L} \frac{\sum_{s_i \in S_i(n^{\ell+1}_i)} \sigma_i(s_i)}{\sum_{s_i \in S_i(n^{\ell}_i)} \sigma_i(s_i)},
\end{eqnarray} (where we take $n_{L + 1} = n$) is the telescopic product for which the numerator of the $\ell$-th term of the product equals to the denominator of the $\ell+1$-th term of the product. Adjacent product terms cancel each other out. Thus,
\begin{eqnarray} \rho(n \mid \beta_i, s_{-i}) & = & \frac{\sum_{s_i \in S_i(n)} \sigma_i(s_i)}{\sum_{s_i \in S_i(n^1_i)} \sigma_i(s_i)}.
\end{eqnarray} Since $n_i^1$ is in player $i$'s first information set on the path towards $n$, we have $S_i(n_1) = S_i$. Hence,
\begin{eqnarray} \rho(n \mid \beta_i, s_{-i}) & = & \frac{\sum_{s_i \in S_i(n)} \sigma_i(s_i)}{\sum_{s_i \in S_i} \sigma_i(s_i)}.
\end{eqnarray} Since trivially any strategy of player $i$ allows her first information set to be reached, we have $\Sigma_{s_i \in S_i} \sigma_i(s_i) = 1$. Thus,
\begin{eqnarray} \rho(n \mid \beta_i, s_{-i}) & = & \frac{\sum_{s_i \in S_i(n)} \sigma_i(s_i)}{1} = \rho(n \mid \sigma_i, s_{-i}).
\end{eqnarray} This completes the proof of the theorem. \hfill $\Box$\\

\begin{figure}[h!]
\caption{An example showing that U1 is required for Kuhn's Theorem \label{U1example}}
\begin{center}
\includegraphics[scale=0.1]{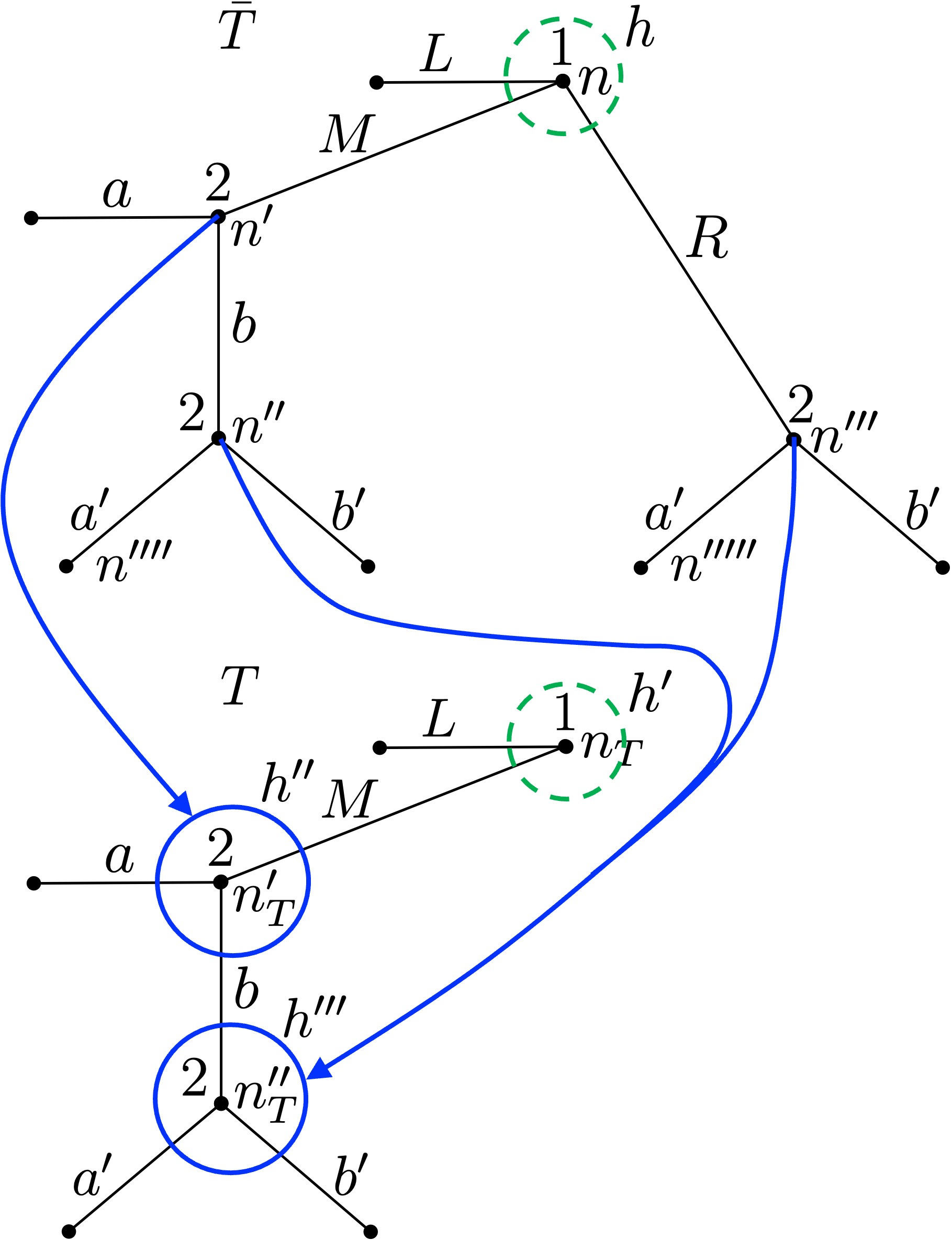}
\end{center}
\end{figure}

There are several subtleties underlying our main result. Of primary interest is condition U1 (Nondelusion), a key restriction whose significance we now expand upon and make explicit. For purposes of comparison, we introduce U1* (Generalized Reflexivity), a more permissive counterpart that has been employed in lieu of U1 in prior work (see Heifetz, Meier, and Schipper, 2013, Schipper, 2021). Thereafter, the example of Figures~\ref{U1example} illustrates that U1 is required for Kuhn’s theorem; by contrast, U1* does not suffice. In past works (see Heifetz, Meier, and Schipper, 2013, Schipper, 2021), the following condition was utilized instead of U1.

\begin{itemize}
\item[U1*] Generalized reflexivity: If $T' \preceq T$, $n \in T$, $h_i(n) \subseteq T'$ and $T'$ contains a copy $n_{T'}$ of $n$, then $n_{T'} \in h_i(n)$.
\end{itemize}

In words, U1* says that if a tree contains the information set of a node and has its copy, then the copy must be in that information set, whereas U1 says that information sets of a node must contain a copy of that node. It is immediate that if a game satisfies U1 then it satisfies U1*, because by U1, for any $h_i (n) \subseteq T'$, it follows that $T' $ contains a copy $n_{T'}$ of $n$ and $n_{T'} \in h_i (n)$. On the other hand, it is not the case that if a game satisfies U1* then it satisfies U1. The example of Figure~\ref{U1example} demonstrates this explicitly. Moreover, the game of Figure~\ref{U1example} satisfies perfect recall but has mixed strategies for which there is no equivalent behavior strategy. Thus, the example shows that U1* is insufficient for Kuhn's Theorem. 

In the example of Figure~\ref{U1example} there are two trees, $\bar{T} \succ T$, and there are two players. Player 1 has two information sets, $h$ in tree $\bar{T}$ and $h'$ in tree $T$ both indicated with green dashed lines. She has three feasible actions in tree $\bar{T}$ at $h$, left, middle and right, and has two feasible actions in tree $T$ at $h'$, left and middle. Player 2 has two information sets, $h''$ and $h'''$ as indicated with blue lines, both of which are in tree $T$. He has two feasible actions at $h''$, $a$ and $b$, and two feasible actions at $h'''$, $a'$ and $b'$. In this example, player 2 is unaware of the right action for player 1 and that node $n'''$ is a possible contingency since his information sets are in $T$. This example satisfies U1*, in particular because $n'_T \in h_2 (n')$ and $n''_T \in h_2(n'')$, but violates U1 because $h_2 (n''')$ does not contain a copy of $n'''$. Notably, Perfect recall (I6) is satisfied because action $b$ for player 2 at node $n'$ leading to $n''$ is the only path in $\bar{T}$ where the action of a player at a node leads to another node where the player is active, and the only node in $h_2 (n'')$ is $n''_T$, action $b$ is played on the path from $n'_T$ to $n''_T$ in $T$, and $h_2 (n') = h_2 (n'_T)$. 

Now, consider a mixed strategy for player 2 consisting of a mixture between $aa'$ and $bb'$ (e.g., probability $\frac{1}{2}$ on $a a'$ and probability $\frac{1}{2}$ on $b b'$). We claim there is no equivalent behavior strategy for this mixed strategy. Observe that with this mixed strategy, $n''$ and $n'''''$ are allowed to be reached with positive probability whereas $n''''$ cannot. For latter, note that $n''''$ can only be reached by playing $b a'$, a strategy assigned probability zero under the mixed strategy. Node $n''$ is reached with positive probability in case $bb'$ is played, which is assigned positive probability under the mixed strategy. Node $n'''''$ is reached with positive probability under strategy $a a'$, which is assigned positive probability under the mixed strategy. But to replicate the distribution of terminal nodes in tree $T$ induced by the mixed strategy, we need the behavior strategy to assign probability $\frac{1}{2}$ to $a$ and $b$ each at information set $h''$ and probability 1 to $b'$ at information $h'''$. But then $n'''''$ is reached with probability zero in tree $\bar{T}$ with this behavior strategy.

This example illustrates some of the difficulties involved in obtaining Kuhn's Theorem beyond standard games. Challenges arise under U1* because a player can mistakenly believe that something has occurred when it has not because she is unaware of what has actually occurred. For instance, in the game of Figure~\ref{U1example}, when player 1 took action $R$ in tree $\bar{T}$, player 2's information set is $h'''$. Thus, player 2 believes that player 1 took action $M$ in tree $T$ because he is unaware of any other action of player 1 that would allow information set $h'''$ to be reached. In fact, he even falsely believes that he must have taken action $b$ at information set $h''$ because otherwise there is no way for him to make sense of arriving at information set $h'''$. In other words, while Perfect recall (I6) rules out that the player forgets an action he has taken, it does not rule out that he makes up having taken an action that in fact he did not take. U1 rules out such delusions about what had occurred. Of course, U1 does not rule out unawareness altogether because the information set at a node in one tree may still consist of nodes in a less expressive tree.

\subsection{Conditions for a Characterization}

While most textbooks in game theory just focus on the arguably more relevant direction of Kuhn's Theorem for standard games without unawareness as Theorem~\ref{Kuhn} does for games with unawareness, Kuhn's (1953) original theorem is a characterization. That is, he proved that in every game of the extensive form without unawareness (in which all decision nodes have at least two actions), the game satisfies perfect recall \emph{if and only if} for every mixed strategy there is an equivalent behavior strategy. The `converse' of Kuhn's Theorem shows the implications for information sets when for each player all causally correlated behaviors can be represented equivalently as products of independent behaviors at different information sets. Given that information sets in games with unawareness not only model the evolution of information but also the evolution of awareness, it is naturally to ask whether in this more complex setting a converse can be obtained under conditions that do not rule out asymmetric awareness altogether. This section contributes towards identifying general conditions under which a characterization is obtained in situations beyond standard games.

\begin{figure}[h!]
\caption{An example showing that U1 is required for the converse of Kuhn's Theorem \label{kuhncounterexample}}
\begin{center}
\includegraphics[scale=0.1]{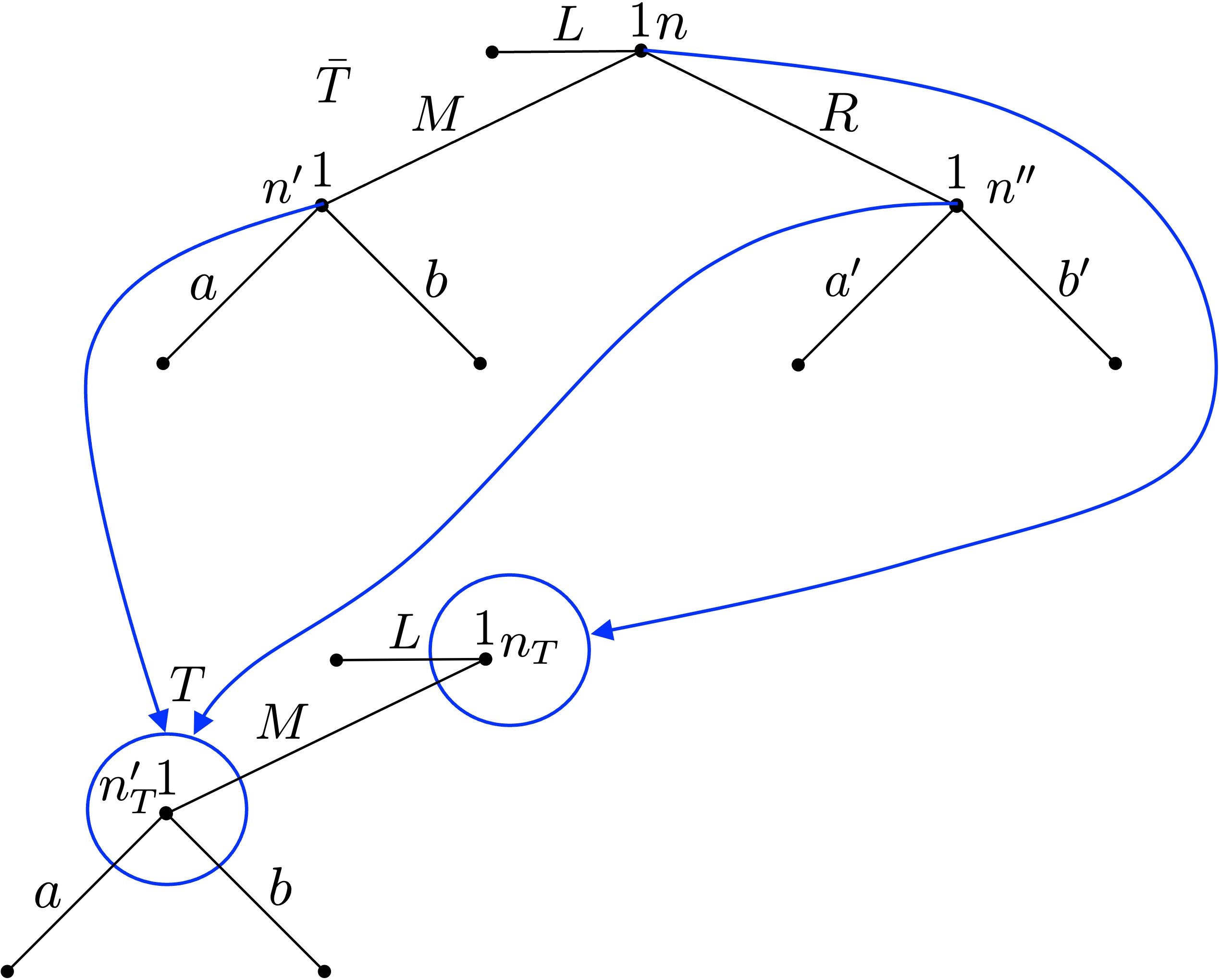}
\end{center}
\end{figure}
First, we show with the example of Figure~\ref{kuhncounterexample} that U1* is also insufficient for obtaining a converse. The adoption of U1 is motivated precisely by the need to preclude situations like in Figures~\ref{U1example} and~\ref{kuhncounterexample}. Figure~\ref{kuhncounterexample} is an example that satisfies U1* but violates U1 and Perfect recall (I6) despite having an equivalent behavior strategy for each mixed strategy. 

There are two trees, $\bar{T} \succ T$ in Figure~\ref{kuhncounterexample}. There is a single player 1 moving at two information sets in this example, both of which are in $T$ as indicated in blue. Since Player 1 does not have information sets in $\bar{T}$, she is never aware of the right action at node $n$. Consequently, we have an equivalence of mixed and behavior strategies. This is because Player $1$ can achieve at $T$ any distribution over terminal nodes with either notion of strategy, and because right is never playable at $h_1 (n)$ (by either notion of strategy) the distribution induced at $\bar{T}$ must be identical to that at $T$. Perfect recall does not hold however, as right is played by Player $1$ on the path from $n$ to $n''$ in $\bar{T}$ but there is no path in $T$ where right is played before $n'_{T}$. Situations like Figure~\ref{kuhncounterexample} where a player can be unaware of her own past action must therefore be precluded for the converse to obtain. U1* is satisfied by the example because $n'_T \in h_1 (n')$ and $n_T \in h_1 (n)$; U1 precludes it because a copy of the node following the right action, $n''$, is not in player 1's state of mind following the right action, $h_1 (n'')$. U1 rules out the intriguing situation of players being unaware of \textit{their own} past action that could lead to another information set of theirs, a situation illustrated in the example of Figure~\ref{kuhncounterexample}.
\begin{figure}[h!]
\caption{An example showing the failure of the converse of Theorem 1 \label{kuhncounterexample2}}
\begin{center}
\includegraphics[scale=0.1]{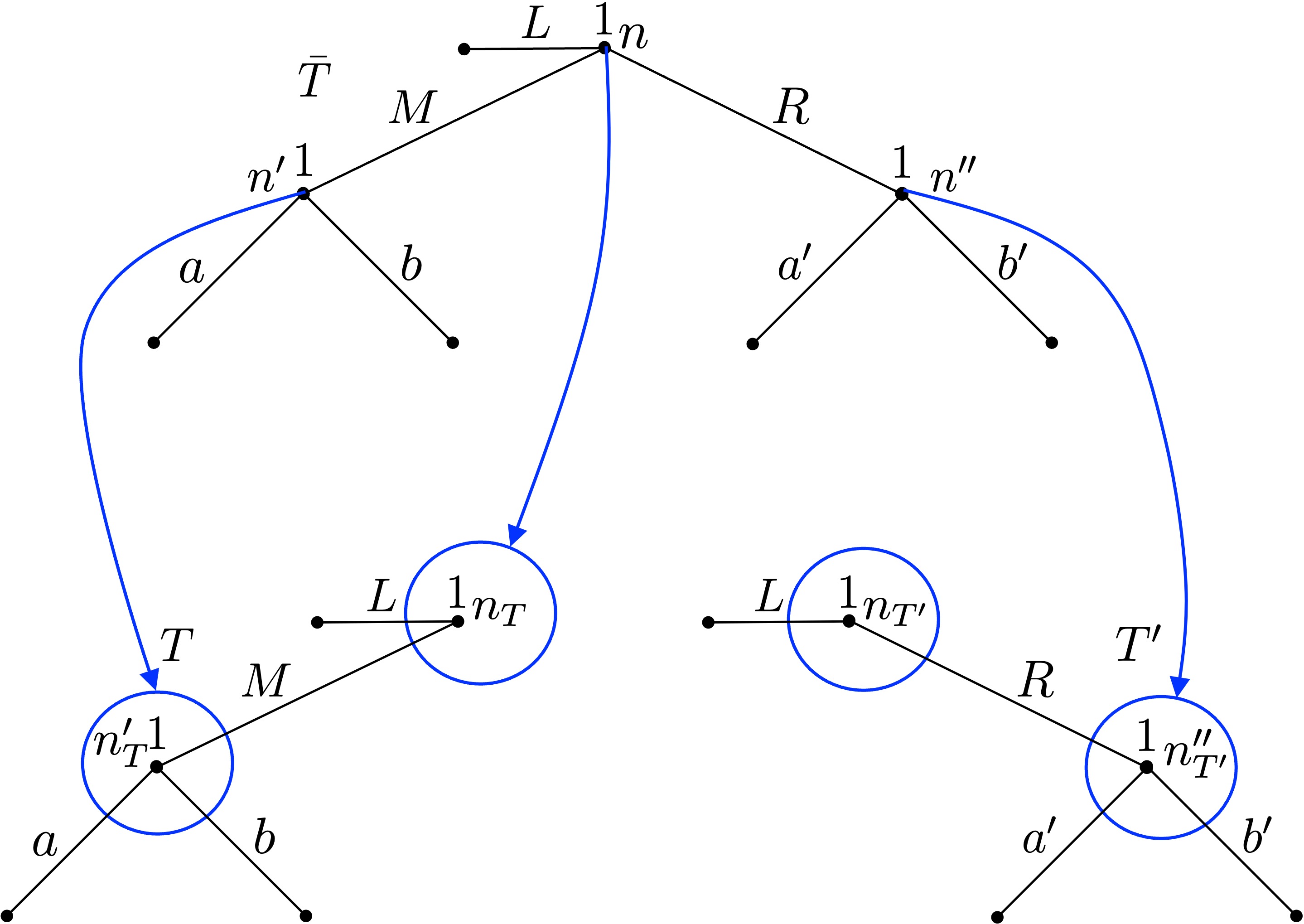}
\end{center}
\end{figure}

Unfortunately, imposing U1 does not yield the converse either as the next example of Figure~\ref{kuhncounterexample2} demonstrates. In Figure~\ref{kuhncounterexample2} there are three trees, $\bar{T}, T $ and $ T'$, where $\bar{T} \succ T$ and $\bar{T} \succ T'$. There is a single player 1 who has four information sets in total indicated in blue, two in tree $T$ and two in tree $T'$. We obtain an equivalence between mixed and behavior strategies because all the information sets are singletons, therefore Player $1$ can achieve at $T$ and $T'$ any distribution over terminal nodes with either notion of strategy, and the distribution induced at $\bar{T}$ must be identical to that at $T$ because $h_1(n) \subseteq T$, $h_1(n') \subseteq T$, and right is not a feasible action for player 1 at $h_1 (n)$. However in order for perfect recall to be fulfilled we require both $h_1 (n) \subseteq T$ and $h_1 (n) \subseteq T'$ which is ruled out by U0.

Note that in games of the extensive form with unawareness, Perfect recall (I6) imposes also a degree of consistency on the structure of information sets and nodes \textit{across} trees. There is one scenario where equivalence between mixed and behavior strategies does not yield this added level of consistency: whenever information sets of the same player at two nodes along the same path are contained in different trees. When this happens, equivalence between mixed and behavior strategies impose fewer structural restrictions on the trees because the information sets are not on the same path and the converse fails precisely due to this weakening. This is what happens in the example of Figure~\ref{kuhncounterexample2} just discussed. 

This motivates the introduction of a new condition, I8, which remedies the aforementioned failure of the converse and together with existing conditions obtains a characterization of perfect recall with the equivalence between mixed and behavior strategies.\footnote{We name this condition I8 rather than I7 because a different condition has been named I7 in earlier works by Schipper (2021).} In words, I8 requires the information sets of any two nodes along the same path where some player is active to be contained in the same tree, thus precluding examples like Figure~\ref{kuhncounterexample2} because there is a path from $n$ to $n''$ but the information set of $n''$ is in $T'$, whereas the information set of $n$ is in $T$.

\begin{itemize}
\item[I8] Maintained awareness along paths: If there is a path $n, \dots , n' \in T$ such that $i \in P(n) \cap P(n')$, then for any $T' \in \mathbf{T}$ such that $ h_i(n) \subseteq T'$, we have that  $h_i(n') \subseteq T'$.\footnote{Another way of stating the I8 condition is the following: For any nodes $n,n'$ in a tree $T$ such that $i \in P(n) \cap P(n')$ and $n \lessdot n'$, we have that $T_{h_i (n)} = T_{h_i (n')}$.} 
\end{itemize}

The I8 condition can be viewed as a stronger version of I3, which also places restrictions on information sets of nodes along the same path. In fact, I3 is implied by I8 and I2.\footnote{Alternatively, I3 is also implied by I8 and U0.} Consider that $n' \in h_i(n) \subseteq T'$ and there is a path $n', \dots , n'' \in T'$ such that $i \in P(n') \cap P(n'')$, then by I2 we have $h_i (n') = h_i (n) $ and by I8 we have that $h_i (n'') \subseteq T'$. Condition I8 `Maintained awareness along paths' stipulates that when players make choices vis-à-vis their state of mind, those choices cannot lead to them discovering or forgetting possible moves for any player. In other words, for every player the level of awareness has to be constant along any path in any tree. Obviously, this is a strong assumption as it restricts the evolution of awareness and we do not suggest in any way that this assumption is justified in every context involving asymmetric awareness among players. But it turns out, the strong assumption allows us to derive the converse of Kuhn's Theorem.  

We say that $\Gamma$ is a \textit{restricted} game of the extensive form with unawareness if it is a game of the extensive form with unawareness that also satisfies property I8, and every node $n \in \mathbf{N}$ has at least two actions (that is, $|A_i (n)| \geq 2, \forall n \in \mathbf{N}, \forall i \in I$). We say that the game satisfies perfect recall if I6 also holds for all players $i \in I$.

We now present Theorem~\ref{strongerpr}, which comes in the form of showing that four statements are equivalent. Kuhn's Theorem and its converse follow from the equivalence of statements 1 and 2. Statement 3 is precisely the perfect recall condition in standard games (for any single tree $T$). Statement 4 generalizes statement 3 (and 1) by allowing the nodes of the paths in question, and also the information set, to be in different trees within the perfect recall condition.

\begin{theo}\label{strongerpr} In a \textbf{restricted} game of the extensive form with unawareness $\Gamma$, the following statements are equivalent:
\begin{itemize} 
\item[1.)] The condition of Perfect recall (I6) is satisfied. 

\item[2.)] For every mixed strategy there is an equivalent behavior strategy.

\item[3.)] For any player $i \in I$, $n_1, n_k \in T \cap \mathbf{D}_i$ for some $T \in \mathbf{T}$ such that $n_1 \lessdot_{a_i} n_k$ and $h_i (n_{k}) \subseteq T$, for all $n \in h_i (n_{k})$ there is some $n'$ such that $n' \lessdot_{a_i} n$ and $h_i (n') = h_i (n_1)$.

\item[4.)] For any player $i \in I$, $n_1, n_k, n'_{k} \in \mathbf{D}_i$ with $h_i (n_k) = h_i (n_{k}')$, then $n_1 \lessdot_{a_i} n_k$ for some $a_i \in A_i (n_1)$ implies $n'_1 \lessdot_{a_i} n'_{k}$ for some $n'_1$ and $h_i (n'_1) = h_i (n_1)$.
\end{itemize} 
\end{theo}

\noindent \textsc{Proof. } ``1.) $\Rightarrow$ 2.)'': Follows from Theorem 1.

``2.) $\Rightarrow$ 3.)'':   Suppose not. That is, we have equivalence between mixed and behavior strategies for every player, but for some player $i \in I$, $n_1, n_k \in T \cap \mathbf{D}_i$ for some $T \in \mathbf{T}$ such that $n_1 \lessdot_{a_i} n_k$ and $h_i (n_{k}) \subseteq T$, there is some $n'_k \in h_i (n_{k})$ such that there is no $n'_1$ such that $n'_1 \lessdot_{a_i} n'_k$ and $h_i (n'_1) = h_i (n_1)$.  Let $a'_i \in A_i (h_i (n_k))$ denote some action available at nodes $n_k$ and $n'_k$, and let $n_{k+1}$ and $n'_{k+1}$ denote the immediate successors of $n_k$ and $n'_k$ respectively following action $a'_i$. Select $s_i \in S_i (n_1 , a_i)\cap S_i (n_k) $ such that $ s_i \notin S_i (n_k , a'_i)$, and some $s'_i \in S_i (n'_k , a'_i)$ such that $s'_i \notin S_i (n_k)$. This is always possible as decision nodes have at least two actions. Construct mixed strategy $\sigma_i \in \Delta S_i $ such that $0 < \sigma_i (s_i) = 1 - \sigma_i (s'_i) < 1$. Then, it follows that $ \rho_i (n_{k+1} \mid \sigma_i) = \Sigma_{s''_i \in S_i (n_{k+1}) } \sigma_i (s_i) = 0$ and $\rho_i (n'_{k+1} \mid \sigma_i)  = \Sigma_{s''_i \in S_i (n'_{k+1}) } \sigma_i (s'_i)  > 0$. Then, for some arbitrary $s_{-i} \in S_{-i} (n_k)$ and $s'_{-i} \in S_{-i} (n'_k)$, we have that $ \rho (n_{k+1} \mid \sigma_i , s_{-i}) = 0$ and $\rho (n'_{k+1} \mid \sigma_i , s'_{-i})  > 0$ and there is no equivalent behavior strategy $\beta_i$ such that $ \rho (n_{k+1} \mid \beta_i , s_{-i}) = 0$ and $\rho (n'_{k+1} \mid \beta_i , s'_{-i})  > 0$ as we would require $\beta_i (h_i (n_k)) (a'_i) = 0$ and $\beta_i (h_i (n'_k)) (a'_i) = \beta_i (h_i (n_k)) (a'_i) > 0$, a contradiction.

``3.) $\Rightarrow$ 4.)'':  Suppose by contradiction that $n_1 \lessdot_{a_i} n_k$ for some ${a_i} \in A_i (n_1)$, however there is some $n'_k$ such that $h_i (n'_{k}) = h_i (n_k)$ but there does not exist $n'_1$ such that $n'_1 \lessdot_{a_i} n'_{k}$ and $h_i (n'_1) = h_i (n_1)$. Let the tree that contains $n_1 , n_k$ be $T$, the tree that contains $n'_k$ be $T'$, and the tree that contains the information set $h_i (n_k)$ be $T''$. This is without loss of generality because $T, T'$ and $T''$ do not have to be distinct trees. By U0, it follows that $T'' \preceq T$, $T'' \preceq T'$. By U1 there is a copy of $n_k$ and $n'_k$ in $h_i (n_k)$, i.e., $(n_k)_{T''}, (n'_k)_{T''} \in h_i (n_k)$. By I8, we have $h_i (n_1) \subseteq T''$ and again by U1 there is a copy of $n_1$ in $h_i (n_1)$, i.e., $(n_1)_{T''} \in h_i (n_1)$. It follows by definition of copies that $(n_1)_{T''} \lessdot_{a_i} (n_k)_{T''}$. Since $h_i (n_k) = h_i (n'_k)$ by supposition, it follows by U1 and I2 that $h_i((n_k)_{T''}) = h_i (n'_k)$. Then, because $(n'_k)_{T''} \in h_i((n_k)_{T''})$, by 3.) there is some $n'$ in tree $T''$ such that $n' \lessdot_{a_i} (n'_k)_{T''}$. Since $T'' \preceq T'$, in tree $T'$ there is $(n')_{T'}$ such that $(n')_{T'} \lessdot_{a_i} n'_k$. In addition, it follows by I8 that $h_i ((n')_{T'}) \subseteq T''$, and by U1, I2, I5 we have $h_i ((n')_{T'}) = h_i ((n')_{T''}) =  h_i ((n_1)_{T''}) = h_i (n_1)$, contradicting our supposition.

``4.) $\Rightarrow$ 1.)'':  Suppose not. Then there are two nodes $n_1$ and $n_k$ with $n_1 \neq n_k$ with the path $n_1, n_2, ..., n_k$ such that at $n_1$ player $i$ takes action $a_i$ along the path, but there is some $n' \in h_i(n_k)$ where there is no node $n_1' \neq n'$ where there is a path $n_1', n_2', ..., n_{\ell}' = n'$, $h_i(n_1') = h_i(n_1)$ and player $i$ takes action $a_i$ at $n_1'$ along the path. But this means that for player $i$, $h_i (n_k) = h_i (n'_{\ell})$ by I2 and $n_1 \lessdot_{a_i} n_k$ but there is no $n'_1$ such that $n'_1 \lessdot_{a_i} n'_{\ell}$ where $h_i (n'_1) = h_i (n_1)$, a contradiction. \hfill $\Box$\\

The characterization of perfect recall by the equivalence of mixed and behavior strategies of Theorem~\ref{strongerpr} is proved under condition I8, restricting the evolution of awareness to be constant along any path in any tree. It does not preclude asymmetric awareness though. Nevertheless, a natural question is whether this condition can be weakened. In particular, in the example of Figure~\ref{kuhncounterexample2}, along the path from $n$ to $n''$ in tree $\bar{T}$, the evolution of player 1's awareness shifts from tree $T$ to $T'$. Those latter two trees are \emph{incomparable} as neither of them is richer than the other. Arguably such an evolution of awareness is of limited relevance because naturally awareness can only increase. One may hope that weakening I8 from constant awareness along any path of play in each tree $T$ to condition DA (briefly introduced in Section~\ref{model}) of (weakly) increasing awareness along any path of play would still allow us to prove a characterization of perfect recall by the equivalence of mixed and behavior strategies. I.e., recall condition DA: For any tree $T \in \mathbf{T}$ and path $n, ..., n'$ in $T$ for which $i \in P(n) \cap P(n')$, require that $h_i(n) \subseteq T'$ and $h_i(n') \subseteq T''$ implies $T'' \succeq T'$. Clearly, this condition weakens I8 from constant awareness along any path of play in $T$ to weakly increasing awareness along any path of play in $T$. It also precludes the counterexample of Figure~\ref{kuhncounterexample2}. However, the hopes of obtaining a converse with this sensible condition is dismissed by the following example. 
\begin{figure}[h!]
\caption{An example showing that Condition I8 cannot be weakened \label{no_weakening}}
\begin{center}
\includegraphics[scale=0.1]{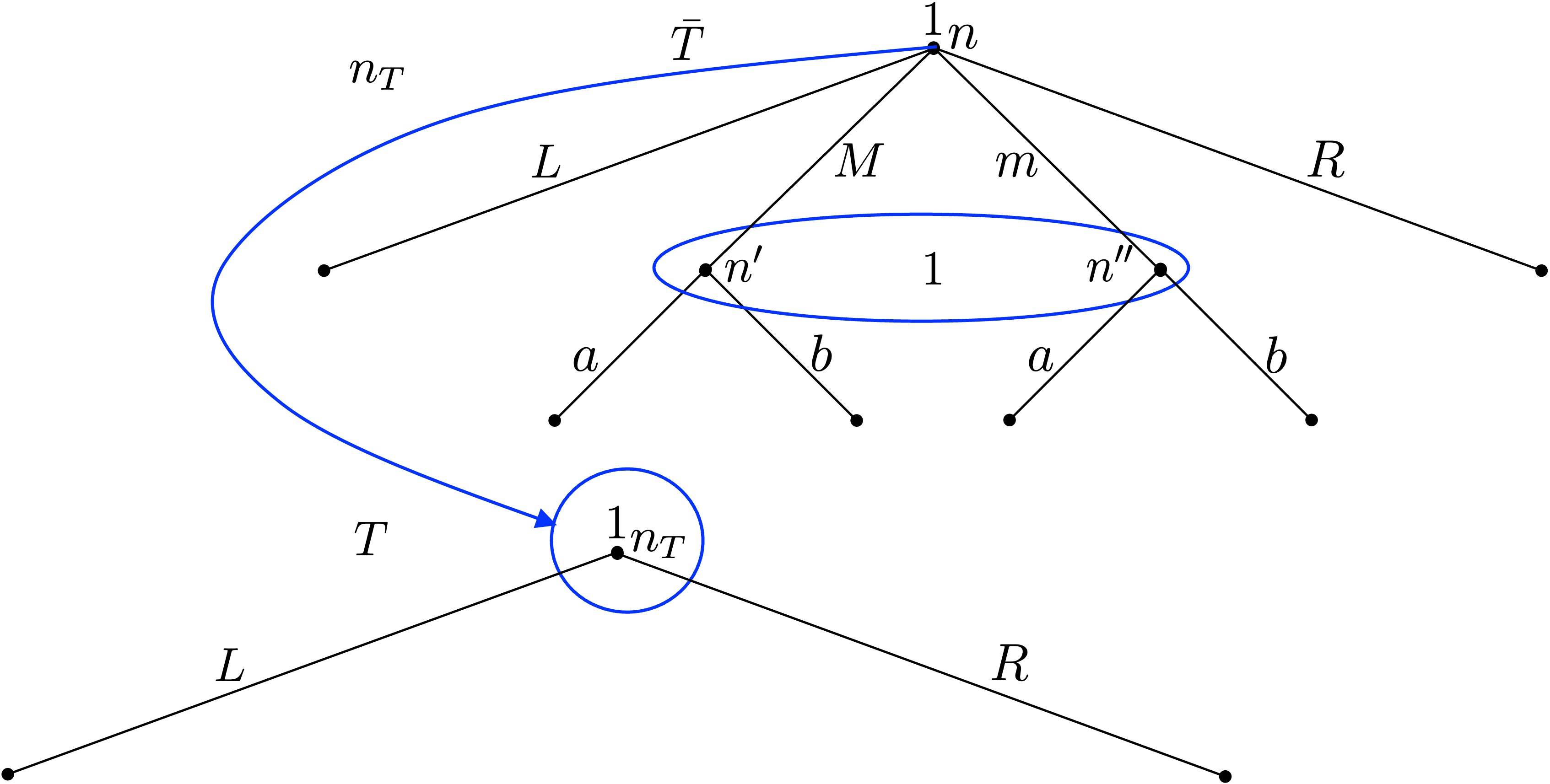}
\end{center}
\end{figure}

In Figure~\ref{no_weakening} there are two trees, the richer tree $\bar{T}$ and the less expressive tree $T$. At $n$ in $\bar{T}$, player 1 is unaware of her actions $M$ and $m$. Thus, her information set at $n$ is in tree $T$ at $n_T$, where she conceives of only two actions, $L$ and $R$. However, on the path from $n$ to $n'$ or from $n$ to $n''$ in tree $\bar{T}$, she would become aware of the existence of both actions $M$ and $m$. She would not know though whether she took action $M$ or $m$, as indicated by her information set at $n'$ and $n''$ in tree $\bar{T}$. This game satisfies all properties except Perfect recall (I6) and I8. On the path from $n$ to $n'$ (and similarly for the path from $n$ to $n''$) in tree $\bar{T}$, her awareness strictly increases. Thus, the example violates I8 but satisfies condition DA. The example violates Perfect recall (I6) because both $n'$ and $n''$ are in the same information set. We clearly have equivalence for mixed and behavior strategies though, because any distribution over terminal nodes can be achieved with either notion of strategy in tree $T$, and there is no feasible action player 1 can take at information set $h_1(n)$ that leads to information set $h_1(n')$; therefore the nodes succeeding $h_1(n')$ can never be reached with positive probability with either notion of strategy. Thus, the example violates the converse. The example demonstrates that essentially condition I8 is the weakest condition under which we are able to obtain the converse. 

Earlier, we observed that in games of the extensive form with unawareness, Perfect recall (I6) imposes also a degree of consistency on the structure of information sets and nodes \textit{across} trees that is not necessarily implied the equivalence between mixed and behavior strategies. Thus, weakening Perfect recall (I6) may present an alternative route to obtaining the converse, at least at a first glance. However, for the converse to obtain, perfect recall must be weakened in a way such that the example of Figure~\ref{kuhncounterexample2} is allowed. In particular, it must allow that once player 1 reaches the information set at node $n''$, she becomes unaware of what she had been aware of previously at $n$, namely the existence of action $M$ followed by actions $a$ or $b$. Losing awareness is an extreme form of forgetting, where the player does not even realize that she has forgotten. In our view, a sensible notion of ``perfect recall'' should not permit this form of extreme forgetting. Accordingly, we believe that trying to obtain the converse in games with unawareness by weakening the Perfect recall (I6) assumption leads us astray from the overall goal. Theorem 2, Part 3.), together with Figure~\ref{no_weakening} essentially demonstrate the weakest setting within our framework in which the assumption of perfect recall for standard games still characterizes the equivalence of mixed and behavior strategies. 

To summarize, from Theorem~\ref{strongerpr} and the examples of Figures~\ref{kuhncounterexample2} and~\ref{no_weakening} we conclude that the converse can be obtained for games with unawareness but only at the cost of restricting the evolution of awareness to constant awareness along paths of play in each tree.

\subsection{Equivalence in Realization of Nodes}

In games of the extensive form with unawareness there are two distinct notions of a strategy profile being consistent with a node. The first notion we introduced already at the end of Section~\ref{strategies} and called it a ``strategy profile reaching a node''. While a player may expect a strategy profile to reach a node, it can be the case in games with unawareness that a different node actually occurs. This is because the player is unaware of actions that a player with more awareness may take (since each player just considers the partial strategies consistent with her awareness level). This begs the question whether strategies that are equivalent with respect to nodes reached are also equivalent with respect to nodes that actually occur. Note that both notions of a node being consistent with a strategy are relevant. The notion of a strategy reaching a node is relevant for extensive-form rationalizability, also called strong rationalizability, due to Pearce (1984) and Battigalli (1997) and extended to games of the extensive form with unawareness by Heifetz, Meier, and Schipper (2013), as well as prudent rationalizability (Heifetz, Meier, and Schipper, 2021, Schipper and Woo, 2019) and versions thereof (i.e., Francetich and Schipper, 2026), whereas the notion of a node occurring with a strategy is crucial for self-confirming equilibrium (Battigalli and Guaitoli, 1997, and Battigalli and Bordoli, 2025) as extended to games of the extensive form with unawareness by Schipper (2021). As we will observe, the notions of node reached and node occurring coincide in games that feature only one tree (as in standard games without unawareness), but can diverge in games with unawareness that have more than one tree. Consequently, our model allows us to make explicit the differences between these two notions, compare them accordingly and illustrate some relationships between the ``subjective'' anticipation of play and the ``objective'' play.

We say that node $n \in \bar{T}$ in the upmost tree $\bar{T}$ \emph{occurs} with strategy profile $s=\left( s_{j}\right)_{j\in I}\in S$ if $n$ is on the path of play in $\bar{T}$ given the players' actions and nature's moves $\left(s_{j}\left( h_{j}(n') \right)\right)_{j \in P(n')}$ in nodes $n' \in \bar{T}$. We extend the notion to any node in any tree by saying that node $n \in T$ occurs with strategy profile $s = \left(s_{j}\right)_{j\in I}\in S$ if there is $n' \in \bar{T}$ s.t. $n'_{T} = n$ and $n'$ occurs with $s$. This is well-defined because $\mathbf{T}$ is a join semi-lattice and therefore there will always be an upmost tree. In particular, for any $T \in \mathbf{T}$ and $n \in T$ there is a node $n' \in \bar{T}$ such that $n'_T = n$. We say that the strategy $s_{i}\in S_{i}$ \emph{allows the node $n$ to occur} if there is a strategy profile $s_{-i}\in S_{-i}$ of the other players (and possibly nature) such that $n$ occurs given the strategy profile $\left( s_{i},s_{-i}\right)$.

We say that information set $h_{i}\in H_{i}$ occurs with strategy profile $s\in S$ if some node $n \in \mathbf{D}_i$ with $h_i(n) = h_i$ occurs with $s$. Note that for this definition we do not require $n \in h_i$. Analogously, we say that the strategy $s_{i}\in S_{i}$ \emph{allows the information set $h_i$ to occur} if there is a strategy profile $s_{-i}\in S_{-i}$ of the other players (and possibly nature) such that $h_i$ occurs given the strategy profile $\left( s_{i},s_{-i}\right)$.
\begin{figure}[h!]
\caption{Illustration of Occur vs. Reached \label{infopath}}
\begin{center}
\includegraphics[scale=0.1]{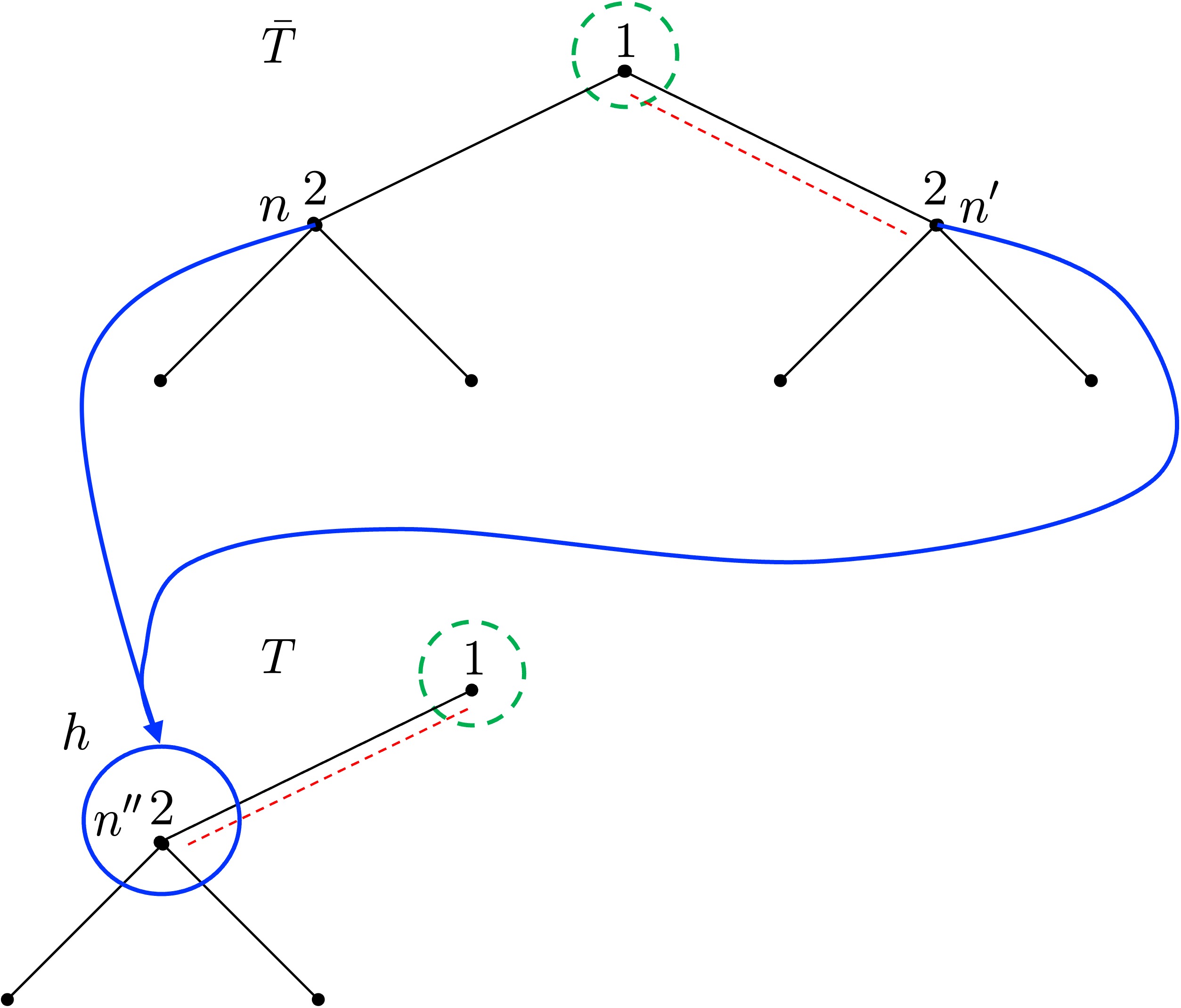}
\end{center}
\end{figure}

The following two examples will help to clarify the definition and its difference to the notion of a strategy reaching a node/information set. Consider first the example in Figure~\ref{infopath}. There are two trees, $\bar{T} \succ T$, and two players, 1 and 2. Player 1 moves first. If she moves left in tree $\bar{T}$, then player 2 remains unaware of her middle action. This is shown in Figure~\ref{infopath} by the blue arrow and disk (i.e., information set $h$ in $T$) upon player 1 moving left. Otherwise, if player 1 moves right in tree $\bar{T}$, player 2 becomes aware of middle (i.e., information set $h''$). (Player 1's initial information sets are indicated by disks with green dashed boundaries.) Consider the strategy of player 1 indicated by the red dashed edges. This strategy allows only nodes $n'''$ and $n'$ to be reached. As player 1 moves first, any strategy profile $s$ of players which includes the above strategy for player 1 will also reach $n'''$ and $n'$. Yet, the nodes that are allowed to occur with this strategy are $n'''$ and $n$. Again, $n'''$ and $n$ will also occur given strategy profile $s$ as player 1 moves first. Thus, this example shows that the nodes reached (resp., a strategy allows to be reached) may differ from nodes occurring (resp., a strategy allows to occur). Note though that they are not disjoint and that this intersection contains a node $n'''$ in the upmost tree $\bar{T}$. In terms of information sets, the strategy profile $s$ reaches only $h'$ but the only information set occurring with this strategy profile is $h$. Thus, the example demonstrates that the information sets reached by a strategy profile may even be disjoint from the information sets occurring.
\begin{figure}[h!]
\caption{Another Illustration of Occur vs. Reached \label{infosetoccur}}
\begin{center}
\includegraphics[scale=0.1]{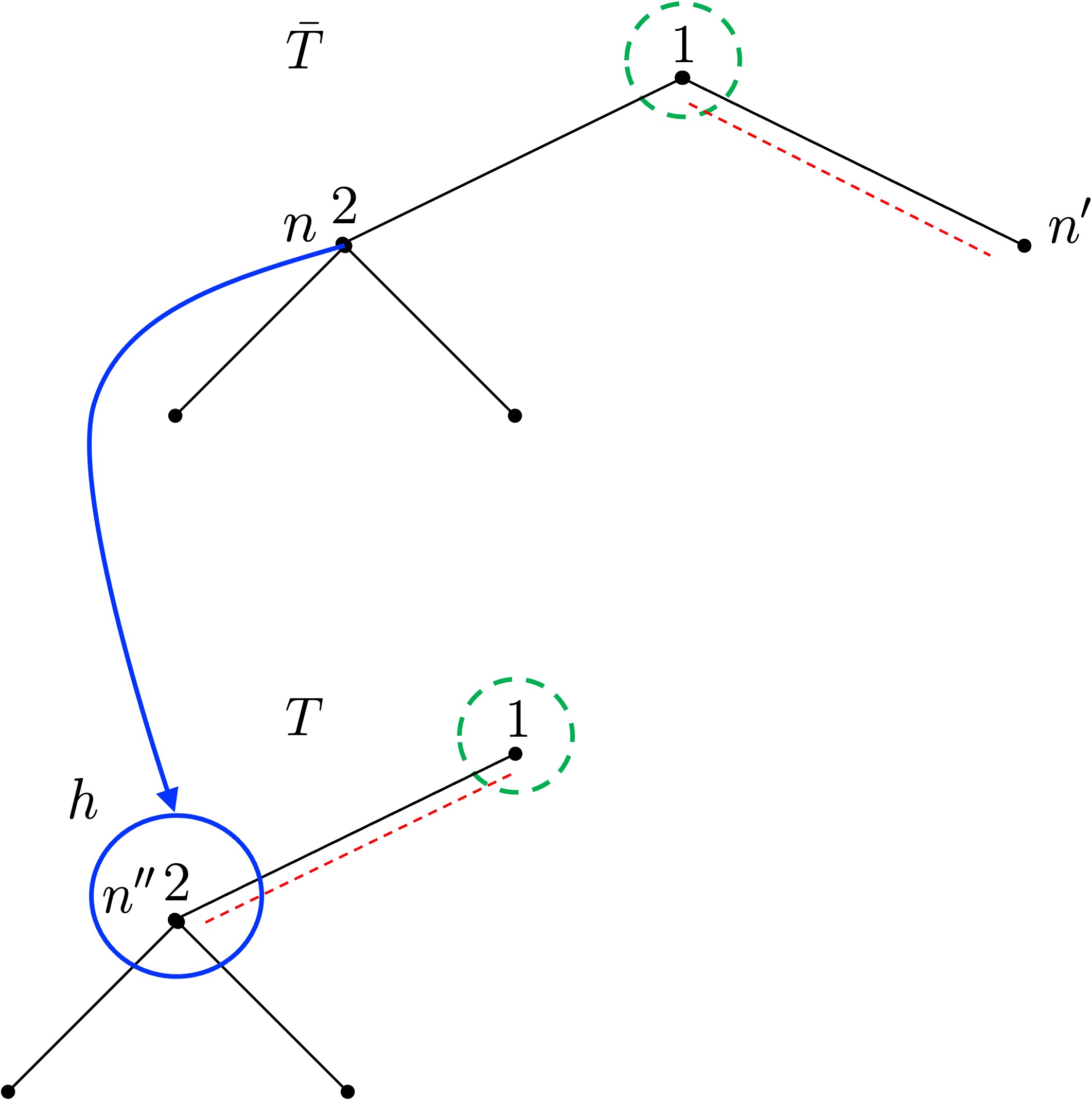}
\end{center}
\end{figure}

The example in Figure~\ref{infopath} has the feature that if an information set is reached (resp., occurs) with a strategy then also a node in this information set is reached (resp., occurs). With respect to the notion of occurring, this may not be the case in general as the next example shows. In Figure~\ref{infosetoccur} there are also two trees, $\bar{T}$ and $T$. Obviously, tree $\bar{T}$ is more expressive than $T$ as it contains also the right action for player 1. Player 1's information sets are again indicated with green dashed boundaries and player 2's information set is indicated in blue. 
Consider the strategy of player 1 indicated by the red dashed edges and let $s$ again denote a strategy profile of players that includes the above strategy of player 1. With respect to nodes, the strategy profile $s$ reaches $n'$ and $n''$. Yet, only $n'$ occurs with this strategy profile. With respect to information set $h$, it is both reached and occurs given strategy profile $s$. Note though that $h$ occurs despite the fact that its only element node $n''$ does not occur with the strategy profile $s$. This is not a defect of the notion of a node/information set occurring but simply reflects the fact that in a game with unawareness the node occurring may not be congruent with what the player believes to occur in her state of mind. Note though that the player is not deluded either as she just misses an important fact rather than ``making things up''.

We summarize the examples:

\begin{rem} In a game of the extensive form with unawareness, if $s$ reaches $n$ then it is not necessarily the case that $n$ occurs with $s$. Similarly, if $n$ occurs with $s$ then it is not necessarily the case that $s$ reaches $n$. Moreover, information set $h$ may occur with a strategy profile $s$ even though $n$ with $h = \{n\}$ does not occur with $s$.
\end{rem}

For the upmost tree $\bar{T}$ the following observation follows directly from the definitions:
\begin{rem}\label{upmost} Consider a game of the extensive form with unawareness with the upmost tree $\bar{T}$. A strategy profile $s$ reaches $n \in \bar{T}$ if and only if $n$ occurs with $s$.
\end{rem}

The observation means that the notions of node reached and node occurring with a strategy really depend on how these notions apply to less expressive trees $T \prec \bar{T}$. The notion of ``reached'' invokes the $T$-partial strategies to determine which node in $T$ is reached and is thus a more subjective notion (from the point of view of a player who considers tree $T$ and lower trees). The notion of ``occur'' invokes actions induced by the strategies in the highest possible tree. Thus it models the ``actual'' or ``objective'' play.

The following corollary follows now directly from the fact that any standard game of the extensive form (i.e., without unawareness) features just one tree.

\begin{cor} In a standard game of the extensive form (i.e., without unawareness), a strategy profile $s$ reaches $n$ if and only if $n$ occurs with $s$.
\end{cor}

We extend the definitions of information sets occurring to behavior and mixed strategies in the obvious way by considering nodes/information sets occurring with strict positive probability.

Let $N(s)$ denote the set of nodes in $\mathbf{N}$ that are reached with strategy profile $s$. Moreover, denote by $O(s)$ the set of nodes in $\mathbf{N}$ that occur with strategy profile $s$. We can now relate the notions of a node being reached and a node occurring with the following lemma:

\begin{lem}\label{consistent} Consider a game of the extensive form with unawareness. For any player $i \in I^0$ and strategies $s_i, s'_i \in S_i$, if $N(s_i, s_{-i}) = N(s'_i, s_{-i})$ then $O(s_i, s_{-i}) = O(s'_i, s_{-i})$ for any $s_{-i} \in S_{-i}$. The converse does not necessarily hold.
\end{lem}

\noindent \textsc{Proof. } For all $s_{-i} \in S_{-i}$, if $N(s_i, s_{-i}) = N(s_i', s_{-i})$ then $N(s_i, s_{-i}) \cap \bar{T} = N(s_i', s_{-i}) \cap \bar{T}$. By Remark~\ref{upmost}
$O(s_i, s_{-i}) \cap \bar{T} = N(s_i, s_{-i}) \cap \bar{T}$ and $O(s'_i, s_{-i}) \cap \bar{T} = N(s_i', s_{-i}) \cap \bar{T}$. Hence, $O(s_i, s_{-i}) \cap \bar{T} = O(s_i', s_{-i}) \cap \bar{T}$. By definition of node occurring, $O(s_i, s_{-i}) = O(s_i', s_{-i})$.

For the converse, we show a counterexample. Consider the game of the extensive form with unawareness in Figure~\ref{infopath}. Further, let $s_1$ ascribe action ``left'' in tree $\bar{T}$ and ``right'' in tree $T$. Moreover, let $s_1'$ ascribe action ``left'' both in tree $\bar{T}$ and $T$. Then information set $h$ occurs both with $s_1$ and $s_1'$. In fact, $O(s_1, s_{-1}) = O(s_1', s_{-1})$ for any $s_1 \in S_1$. Yet, only strategy $s_1'$ reaches $h$ in $T$ while $s_1$ reaches $h'$ in $T$.\hfill $\Box$\\

For any node $n$, any player $i \in I^0$, and any opponents' profile of strategies $s_{-i}$ (including nature if any), let $o(n \mid \beta_i, s_{-i})$ and $o(n \mid \sigma_i, s_{-i})$ denote the probability that node $n$ occurs with $(\beta_i, s_{-i})$ and $(\sigma_i, s_{-i})$, respectively.

\begin{rem} In a game of the extensive form with unawareness, it is not necessarily the case that for each tree $T \in \mathbf{T}$, $o( \cdot \mid \sigma_i, s_{-i})$ defines a distribution over terminal nodes $Z \subseteq T$. E.g., in the example of Figure~\ref{infosetoccur}, no terminal node of $T$ occurs with the strategy of player 1 indicated by the red dashed line. Thus, $o( \cdot \mid \sigma_i, s_{-i})$ may not only be subadditive but may even assign zero to the set of all terminal nodes in a given game tree. It is however always a probability distribution over nodes in the upmost tree $\bar{T}$.
\end{rem}

We use the notion of a node occurring to define another notion of equivalence between strategy that we dub realization-equivalent. For any player $i \in I^0$, a mixed strategy $\sigma_i$ and a behavior strategy $\beta_i$ are \emph{realization-equivalent} if for every profile of opponents' strategies $s_{-i} \in S_{-i}$ and every node $n \in \mathbf{N}$ of the game of the extensive form with unawareness $o(n \mid \sigma_i, s_{-i}) = o(n \mid \beta_i, s_{-i})$. Since information sets can be viewed as functions of nodes, realization-equivalent strategies are also realization-equivalent with respect to the probability of information sets occurring. This is relevant because information sets model also the player's state of mind. We want to assure that strategies are also equivalent with respect to the states of mind that may arise along the play.
\begin{rem} If two strategies are realization-equivalent then the same information sets occur with the same probabilities with both strategies.
\end{rem}

\begin{lem}\label{realequi} In any game of the extensive form with unawareness and perfect recall, if $\sigma_i$ and $\beta_i$ are equivalent to each other, then they are also realization-equivalent.
\end{lem}

\noindent \textsc{Proof.} For any $\sigma_i \in \Delta(S_i)$, $\beta_i \in B_i$, $s_{-i} \in S_{-i}$, $T \in \mathbf{T}$, $n \in T$, $o(n \mid \sigma_i, s_{-i}) = \rho(n' \mid \sigma_i, s_{-i})$ and $o(n \mid \beta_i, s_{-i}) = \rho(n' \mid \beta_i, s_{-i})$ for $n' \in \bar{T}$ such that $(n')_T = n$. Let $\beta_i$ be equivalent to $\sigma_i$. Then the conclusion follows from Theorem~\ref{Kuhn}. \hfill $\Box$\\

Theorem~\ref{Kuhn} and Lemma~\ref{realequi} now imply immediately the following corollary:

\begin{cor}\label{Kuhnrealequi} In every game of the extensive form with unawareness, if player $i$ has perfect recall, then for every mixed strategy of player $i$ there exists a realization-equivalent behavior strategy.
\end{cor}

\subsection{T-Partial Games and T-Partial Strategies}

We return to the fact that in games of the extensive form with unawareness, strategies may not only be an object of choice for a player, but also an object of belief of other players.

Let $S^T_i(n)$ be the set of all $T$-partial strategies of player $i$ that reach $n$, where one should note that $T_n \preceq T$. That is, $s_i \in S^T_i(n)$ where $T \succeq T_n$ if and only if there exists $s_{-i} \in S^T_{-i}$ such that $n$ is on the path of play in $T_n$ given the profile of players' actions and nature's moves $(s_j^{T_n}(h_j(n')))_{j \in P(n')}$ in nodes $n' \in T_n$.

Note that a $T$-partial game is a game of the extensive form with unawareness in which the join of the join-semilattice of trees is $T$. Thus, Lemma~\ref{Kuhnlemma} implies immediately the following corollary:

\begin{cor}\label{Kuhnlemma2} Consider a game of the extensive form with unawareness $\Gamma$. If $\Gamma$ satisfies perfect recall (i.e., I6), then for any player $i \in I$, $n \in N$ with $h_i(n) \in H_i$, and $n' \in h_i(n)$, $S^T_i(n) = S^T_i(n')$ for any $T \succeq T_n$.
\end{cor}

This corollary is relevant because we can view strategies of a player as objects of beliefs of other players. Yet, their beliefs are bounded by their awareness. That is, if player $i$ arrives at information set $h_i$, then her awareness level is given by $T_{h_i}$, the tree that contains information set $h_i$. Thus, she forms beliefs about player $j$'s $T_{h_i}$-partial strategies.

Theorem~\ref{Kuhn} implies now immediately the version for $T$-partial strategies.

\begin{cor} In every game of the extensive form with unawareness, if player $i$ has perfect recall, then for every $T$-partial mixed strategy of player $i$ there exists an equivalent $T$-partial behavior strategy, for $T \in \mathbf{T}$.
\end{cor}

The statement of the original Kuhn's Theorem is now a corollary for $T$ being a least expressive tree or $\mathbf{T}$ being singleton.

It is possible to also define a notion of node occurring with a $T$-partial strategy profile. Yet, such a definition is not very meaningful as the notion of a node occurring aims to characterize nodes that actually (or ``objectively'') occur. To determine such nodes, it is crucial to consider which nodes are reached in the upmost tree $\bar{T}$. $T$-partial strategies, with $T \prec \bar{T}$, are by definition silent on it. Yet, every $T$-partial strategy can be extended to a strategy on the entire join-semilattice of trees $\mathbf{T}$. Such an extension is typically not unique. The nodes occurring will then depend which extension is considered.

\section{Discussion\label{discussion}} 

In games with unawareness, a player may also be unaware of some of her own actions. Our notion of strategy and properties on information sets restrict each player to only take actions of which she is aware. If we were instead to allow a player to take an action that she is unaware of without her realizing that she took such an action, then she could not know and remember such a choice, hence violating perfect recall. This is exactly what happens in the example of Figure~\ref{kuhncounterexample}. The rules of physical moves of the game allow the player to go right initially even though she does not realize that she could go right. Moreover, even after going right, she does not realize that she went right leading her mistakenly to believe she went left. In contrast to standard games without unawareness, in games with unawareness the set of nodes where the player has the same state of mind can be larger than the set of nodes that she considers possible. This gives rise to situations where at her information set the player is unaware of some previous action of her own that allows the node at which she has this information set. The perfect recall assumption rules this out. That is, at an information set the player cannot be unaware of her prior \emph{own} action that allows a node at which she has this information set, ruling out scenarios such as in Figure~\ref{kuhncounterexample}. Note that this example only violates the memory condition for actions. It does satisfy the memory condition for information sets, i.e., at the latter information set in $T$, the player remembers everything she knew at the initial information set at $T$ (all the while being completely unaware of the right action at $n$).

It is natural to explore what would happen if a player is allowed to play an action that he is unaware of. In Figure~\ref{kuhncounterexample}, equivalence between mixed and behavior strategies is only obtained due to right not being playable at $h_1 (n)$. If within our notions of mixed and behavior strategies we allow the player to play right (despite being unaware of it) then equivalence would no longer hold as mixed strategies would be able to achieve a larger set of outcomes than behavior strategies.\footnote{For example, consider the mixed strategy that assigns probability $\frac{1}{2}$ to $(left , left)$ and probability $\frac{1}{2}$ to $(right , right)$, where the first entry in the profiles refers to the action chosen at the initial information set and the second entry refers to the action chosen at the latter information set. There is no equivalent behavior strategy that achieves the distribution over outcomes in $\bar{T}$.} In the sense that the tree $\bar{T}$ models ``actual'' or ``objective'' play, one possible way to look at this question is whether we should allow the actual play and intended play to diverge.\footnote{See Battigalli and De Vito (2021) for recent work distinguishing between intended play and actual play in standard games without unawareness.} If we think of it in relation to real life situations, it certainly does not seem unreasonable to allow this. For instance, a fighter pilot during a stressful dogfight while intending to press one button in a complex cockpit may accidentally press another button he does not even consider at that moment and may entirely fail to realize that he has pressed the wrong button. In day to day life it is also not uncommon for us to make a mistake we cannot even conceive of at the time of acting, for example making typos in the essay we are writing due to pressing the wrong keys and unknowingly invoking shortcuts or commands in the software, all the while remaining completely unaware of the error. Capturing the above situations within our model is an interesting avenue for future research as it would allow us to analyze problems of \emph{situation awareness}, the study of the human factor in complex dynamic environments such as military interactions (e.g., Endsley and Garland, 2000). However, to the extent that we aim to model rational play as the benchmark for real behavior, we want to rule out that players choose to play actions of which they are unaware of. After all, intentional choice can only be done among options that one can conceive of. And by restricting actions to only what is available to a player at an information set, we naturally rule out situations where the player plays actions she is not aware of.


\end{document}